\documentclass[11pt, a4paper]{article}
\usepackage{amsmath, amssymb, mathabx}
\usepackage{graphicx}
\PassOptionsToPackage{hyphens}{url}
\usepackage{hyperref}
\usepackage{apacite}

\title{Toward a formal theory for computing machines made out of
whatever physics offers: extended version}

\author{Herbert Jaeger\thanks{Bernoulli Institute and Groningen
    Cognitive Systems and Materials Center (CogniGron), University of
    Groningen, 9700 AB Groningen, Netherlands}, Beatriz
  Noheda\thanks{Zernike Insitute for Advanced Materials and Groningen
    Cognitive Systems and Materials Center (CogniGron), University of
    Groningen, 9700 AB Groningen, Netherlands}, Wilfred G. van der
  Wiel\thanks{BRAINS Center for Brain-Inspired Nano Systems, MESA+
    Institute for Nanotechnology, University of Twente, 7500 AE
    Enschede, Netherlands and Institute of Physics, Westf\"{a}lische
    Wilhelms-Universität M\"{u}nster, Germany} }

\makeatletter
\let\@fnsymbol\@arabic
\def\@setthanks{\vspace{-\baselineskip}\def\thanks##1{\@par##1\@addpunct.}\thankses}
\makeatother

\begin{document}

\maketitle

\begin{abstract}

  Approaching limitations of digital computing technologies have
  spurred research in neuromorphic and other unconventional approaches
  to computing. Here we argue that if we want to systematically
  engineer computing systems that are based on unconventional physical
  effects, we need guidance from a formal theory that is different
  from the symbolic-algorithmic theory of today's computer science
  textbooks.  We propose a general strategy for developing such a
  theory, and within that general view, a specific approach that we
  call \emph{fluent computing}. In contrast to Turing, who modeled
  computing processes from a top-down perspective as symbolic
  reasoning, we adopt the scientific paradigm of physics and model
  physical computing systems bottom-up by formalizing what can
  ultimately be measured in any physical substrate. This
  leads to an understanding of computing as the structuring of
  processes, while classical models of computing systems describe the
  processing of structures.

  This is a greatly extended version of a perspective article
  \cite{JaegerNohedaVdWiel23a} that appeared in Nature
  Communications.  
\end{abstract}

\newpage
\setcounter{tocdepth}{2}
\tableofcontents

\newpage

\section{Introduction}

The all-overturning powers of digital computing (DC) technologies need
no elaboration. Since a decade or so it has however becoming
increasingly clear that DC technologies are accelerating into a
narrowing lane with regards to energy footprint \cite{AndraeEdler15};
toxic waste \cite{WorldEconomicForum19}; physical, technological and
economical limits of miniaturization \cite{Waldrop16} and
vulnerabilites of ever growing software complexity
\cite{Ebert18}. These conditions have spurred explorations of
alternatives to digital computing.  Currently the most widely and
deeply explored non-digital route to computing is \emph{neuromorphic}
computing \cite{Mead90} --- use biological brains as role model for
energy-efficient and high-throughput parallel algorithms and novel
kinds of microchips. We also see a reinvigorated study of other
unconventional computing paradigms, of which there are
many. They have been introduced under names like \emph{natural
  computing, in-materio computing (or in-materia computing
  \cite{Ricciardi22}), emergent computation, physical computing,
  reservoir computing} \cite{EuropeanCommission09, Adamatzky17ab,
  Jaeger21a}, and they search for computational exploits in a wide
variety of biological, chemical and physical systems and
substrates. Examples are analog electronic computers
\cite{BournezPouly21}, slime moulds \cite{Adamatzky18}, physical
reservoir systems \cite{Tanakaetal18}, DNA reactors
\cite{vanNoortetal02, Doty12}, chemical reaction networks
\cite{Montietal17}, ant colonies \cite{DorigoGambardella97}, or social
decision making networks \cite{Minsky85, McPhailPowersTucker92}. Some
of these initiatives can look back on a long history.

Today a large variety of systems are being investigated in the wide
fields of neuromorphic and other unconventional computing
researches. These systems are artifical or natural, exist as formal
models, digital simulations, manufactured hardware, or are identified
in natural hosts like DNA soups, immune systems, cells, brains or
animal societies. All of these systems 'compute' or ``process
information'' in one way or another. They serve different purposes
like signal processing and control, creative problem solving,
optimization, autonomous decision-making and agent intelligence. Their
behaviour can be shaped (or not) by users according to various
pardigms, including programming, system hardware configuration,
training, evolutionary optimization, or self-organized task
adapation. Physical materials and devices offer a limitless
reservoir of physical phenomena for building unconventional computing
machines.  In turn, these phenomena can be modeled by a likewise
almost limitless range of mathematical constructs. Often these
constructs are quite generic and can be found in almost every
sufficiently complex physical or neural system --- for instance
oscillations, chaos and other attractor-like phenomena; hysteresis;
many sorts of bifurcations and input-induced transits between basins
of attraction; spatiotemporal pattern formation; intrinsic noise;
phase transitions. The pertinent literature for each of them is so
extensive and diverse that it defies a systematic survey. Other
mathematical constructs are more specific, for instance heteroclinic
channels and attractor relics \cite{Rabinovich08,Gros09},
self-organized criticality \cite{Chialvo10,Stiegetal12,BeggsTimme12} or
solitons and waves \cite{LinsSchoener14,Grollieretal20}.

Across the diversity of materials, methods and motives we perceive a
growing awareness (or wish) that there is (or should be) a common
ground from which these diverse branches of research arise, and in
which they can (or might) become re-united --- a unified science of
information processing systems which is more general than, or just
different from, today's canonical science of symbolic-discrete
computing. While this is a vague and distant goal, the relevant
communities are making increasingly energetic efforts to move closer
together, such that they can learn from each other. This is witnessed
by high-profile target articles \cite{SchumanEtAl22,MehonicKenyon22},
interdisciplinary collection volumes \cite{Adamatzky17ab}, conferences
and workshops, large-scale public-funded research projects (some in
the acknowledgements at the end of this article), or newly founded
academic study programs and research institutes (some are listed by
\citeA{MehonicKenyon22}).

At present, most of these activities label themselves as
'neuromorphic'. We see several reasons for the
current prominence of the 'neuromorphic' paradigm: the blazing
achievements of deep neural networks in machine learning; concrete 
technological promises of memristive synapses for in-memory computing;
and the unique standing of brains as the role model which, among all
natural 'computing' systems, is the most complex, powerful and
intriguing one.

We do not want to separate neuromorphic from other unconventional
approaches to 'computing'. Both lines of study can be seen as
belonging together in that they are interested in 'natural' aspects of
computing systems like self-organization, adaptability, learning,
creativity, energy efficiency, noise robustness, error tolerance and
graceful degradation, autonomy, continuous-time interaction with an
environment, or statistical dynamics in large ensembles --- all of
these are not natively connected with the digital-symbolic computing
paradigm.

To preclude misunderstandings we mention that we consider quantum
computing in its classical form --- carefully stabilized qbit carriers
exploiting quantum state superposition for parallel search --- as a
variant of classical symbolic computing rather than as an example of
unconventional computing. The theory and intended applications of
traditional quantum computing are couched in the classical Turing
paradigm, offering faster algorithmic solutions
for tasks that could likewise be solved by Turing machines.

Progress in neuromorphic and other unconventional computing is
slow. While a wealth of ideas, methods, materials, devices,
proof-of-principle demonstrators, and analyses are being generated,
these results remain largely separated by disciplinary boundaries
despite all efforts for community-building. We believe that this state
of affairs will persist as long as there is no unifying formal theory
that could connect the dots. Such a formal theory would be crucial for
a scientific discipline of engineering neuromorphic and other
unconventional computing systems in a principled, systematic way. We
are certainly not the only ones to deplore the absence of such a
foundation in a unifying theory:
\emph{``The ultimate goal would be a unified domain of all forms of
  computation, in as far as is possible...''}
\cite{EuropeanCommission09}; \emph{``As the domain of computer science
  grows, as one computational model no longer fits all, its true
  nature is being revealed... New computers could inform new
  computational theories, and those theories could then help us
  understand the physical world around us''} \cite{Horsman17};
\emph{``there is still a gap in defining abstractions for using
  neuromorphic computers more broadly''} \cite{SchumanEtAl22};
\emph{``The neuromorphic community ... lacks a focus. [...] We need
  holistic and concurrent design across the whole stack [...] to
  ensure as full an integration of bio-inspired principles into
  hardware as possible''} \cite{MehonicKenyon22}.

There already exists a broad spectrum of formal theories that may be
candidates or starting points for a unified theory of neuromorphic and
unconventional computing systems. These theories have been developed
in computer science, theoretical natural sciences, systems engineering
and complex systems research for a variety of goals: to enable an
interpretation of natural processes as 'computing'; to unify the laws
of physics in a concept of 'information'; to help describing and
understanding neural and cognitive processes; to describe complex
engineered or natural systems through conceptual and/or procedural
hierarchies; or to guide the design of computing machines other than
digital-symbolic ones. We highlight the range of existing formal
frameworks by listing some of them --- to underline the confusing wealth and diversity of our findings, we do this in random order:

\begin{itemize}
\item The classical models of \emph{analog computing} systems
  formalize analog mechanical or electronic devices that realize
  real-valued elementary operations like addition or integration can
  be combined in complex system for realizing a hierarchy of
  real-valued functions. These hierarchies had originally been shaped
  in the molds of symbolic-logical theories of Turing-computable
  functions \cite{Shannon41, Moore96}, but later the perspective has
  broadened a lot (surveyed by \citeA{BournezPouly21}).
\item A traditional subfield of AI, \emph{qualitative physics},
  \cite{Forbus88} (closely related: \emph{naive physics},
  \emph{qualitative reasoning}) explores logic-based formalisms which
  capture the everyday reasoning of humans about their mesoscale
  physical environment.

\item Ulf Grenander's \emph{pattern theory}, especially in the
  transparent workout of David Mumford \cite{Mumford94}, offers a
  thoroughly formal account of how (primarily spatial / visual)
  ``patterns'' can be generated, compounded, transformed and
  encoded. Pattern theory is sophisticated --- David Mumford is a
  recipient of the Fields Medal, and he considers pattern theory a
  candidate for ``a mathematical theory underlying intelligence''
  \cite{Mumford02}.
  
\item Insights gained in the fields of \emph{emergent computation}
  \cite{Forrest1990a} steer attention to the powers of collective
  phenomena in dissipative systems, where macrolevel phenomena
  ``self-organize'' from the interactions of microlevel components.

\item \emph{Complex systems modeling general.} Formal models of
  complex natural systems are regularly chiseled out in formats that
  have structural and procedural similarities with formalisms in AI or
  computer science. Such models admit interpretations of structures
  and processes in 'computational' or 'cognitive' or 'information
  processing' terms. A survey cannot be attempted. An example are
  models in motion science, which capture bodily motion patterns of
  animals and humans. Important work in this field views complex,
  continuous physical motion patterns in ways that have strong
  analogies with cognitive and computational processes, by defining
  criteria for segmenting and composing bodily motions hierarchically
  in multiple spatial and temporal scales, modeling their planning and
  execution control, and analysing how they can be semantically
  interpreted by observers \cite{HoganFlash87, ThoroughmanShadmehr00,
    Roetheretal10,Landetal13}.
  
\item The theory of \emph{autopoeitic systems}, established by
  Humberto Maturana and Francisco J.\ Varela \cite{Varelaetal74,
    MaturanaVarelaEnglish84}, explains the stability of biological
  organisms through internal feedback loops in which all parts and
  functions engage in a concert to reproduce themselves. In this
  light, cognitive processes are not seen as based on representations
  of external reality, but as constituting their own inner reality in
  network of interconnected processes. This view challenges
  commonsense and philosophical conceptions of cognitive
  representations, exerting a lasting impact outside theoretical
  biology in epistemology, cognitive science, sociology and other
  fields \cite{Razeto-Barry12}. Principles from autopoeisis have
  frequently been invoked by proponents of behavior-based robotics
  \cite{Brooks91a} and 'new AI' \cite{PfeiferScheier98} to explain how
  intelligent information processing does not need explicit internal
  representations (in the classical AI spirit) of an agent's
  environment.
  
\item \emph{Stream automata} \cite{Endrullisetal20} aim at extending
  the classical theory of finite-state automata to infinite data
  stream processing. This can be seen as a step toward modeling neural
  processing with tools that grow out of classical computer science,
  because brains (and other natural systems that have been regarded as
  processing information) are also stream processing systems.
  
\item \emph{Process calculi} and other mathematical models, including
  the well-known \emph{Petri nets} \cite{PetriReisig08}, aim at
  modeling distributed information processing which unfolds in
  concurrent subprocesses. These formalisms belong to classical
  computer science. A category-theoretical unification is proposed by
  \citeA{WinskelNielsen93}. These modeling tools have been adapted
  outside computer science to model processing of information or
  materials in other engineered or natural systems, notably by Luca
  Cardelli who tailored these tools in many ways to formalize processes in
  biological and chemical systems (example: \emph{membrane computing}
  \cite{Cardelli05, Paun10}).

\item \emph{Interactive symbolic computing.} In some non-standard
  use-cases considered in modern computer science, 'computing' is seen
  as an interaction sequence between a (otherwise classical
  symbolic-discrete) computer and a user. The information feedback
  through the user extends the class of problems that can be solved by
  such interaction pairs beyond the Turing-computable problems
  \cite{WegnerGoldin03}.
  
\item In computer science and systems engineering, \emph{hybrid
    systems} are systems that combine computational and physical
  processing, or software and hardware, or discrete and continuous
  states. Formalisms for modeling such systems are likewise hybrids of
  classical discrete-symbolic models of computer science (often
  automata models) with continuous-state physical modeling inserts
  \cite{Lynchetal03,Geuversetal10}.

\item \emph{Complexity theory for neural networks.}
  \citeA{KwisthoutDonselaar20} consider Turing machines, which upon
  presentation of a task input automatically construct a formal model
  of a spiking neural network that can process this task, and
  investigates the combined consumption of computational resources for
  such twin systems. For the neural network model he allows
  unconventional resource categories like the number of used
  spikes. This work renders spiking neural networks accessible to
  classical theory of computational complexity, but does not specify
  how the neural networks spawned by the Turing machine are concretely
  designed, and the approach is only applicable to a specific formal
  model of neural networks, not to general physical computing systems.

\item \emph{Recurrent neural networks} (RNNs) are neural networks
  whose cyclic connection topology makes them dynamical
  systems. Always present since the beginnings of the study of neural
  networks \cite{McCullochPitts43}, this family of models has risen to
  new levels of importance through their wide use in deep
  learning. Very recently there even seems to be a new surge of
  interest because innovative RNN architectures have become at least
  competitive with, and sometimes superior to transformer networks,
  which at present are deemed the most powerful deep learning systems
  \cite{Zucchetetal23}. The family of RNN models at large is so
  diverse that we cannot attempt to discuss them here in fuller scope.

\item \emph{Reservoir computing} is a special RNN design for supervised
  learning that originated in machine learning \cite{Jaeger01a} and
  computational neuroscience \cite{MaassETAL01a}. A randomly
  connected recurrent neural network is excited by an input signal,
  and from the richly varied nonlinear response signals inside the
  'reservoir' network a trainable output signal is linearly
  combined. This sort of system has been thoroughly investigated by
  mathematicians, revealing which input-output signal transformations
  can be realized \cite{GrigoryevaOrtega18} --- namely the class of
  \emph{fading memory} tasks. We mention reservoir computing here
  separately from RNNs in general because reservoir systems have
  become variously adopted by materials scientists, who replace the
  neural reservoir by nonlinearly excitable physical substrates
  \cite{Tanakaetal18}.
  
\item The Neural Engineering Framework
  \cite{StewartBekolayEliasmith11}, originally developed by Chris
  Eliasmith and Charles Anderson and used in a sizeable community of
  cognitive neuroscientists
  \cite{Bekolayetal14,Eliasmithetal12,Neckaretal19,Taatgen19,Angelidisetal21},
  provides mathematical analyses and design rules for interacting
  modules of spiking neural networks that realize signal processing
  filters, which are specified by ordinary differential equations.

  \item \emph{Cognitive information processing as statistical inference.}
  Intelligent agents that operate in stochastic environments must be
  able to compute probability distributions of the expected
  consequences of their actions. This core idea of a \emph{predictive
    brain} \cite{Clark13} has been formally worked out in numerous
  formats and communities. The mechanical operations needed to reason
  with and about probability distributions are often realized through
  stochastic sampling dynamics \cite{Jaeger21}. Some examples of
   workouts in this spirit:

  \begin{itemize}
  \item The \emph{free energy principle} of \cite{Friston10} casts
    the learning and adaptation of autonomous agents in challenging
    environments in a mathematical formalism that originated in
    statistical physics, information theory and Bayesian
    statistics. Specifically, the efforts of the learner, who needs to
    distil a useful representation of the results of its actions in a
    stochastic environment, is interpreted in terms of minimizing a
    quantity that is formally analogue to free energy in statistical
    physics. The high abstraction level of this mathematical framework
    admits a unified view on a variety of existing models of cognitive
    functions, representations, and learning, which are stated on
    behavioral, cognitive, or physiological levels.
  \item \emph{Hopfield networks} \cite{Hopfield82}, \emph{Boltzmann
      machines} \cite{Ackleyetal85} and the \emph{restricted
      Boltzmann machine} \cite{HintonSalakhutdinov06} are classical
    instances of calling upon methods of statistical physics (in
    particular the Boltzmann distribution) for stochastic neural network
    models. A unifying review is given
    by \citeA{MarulloAgliari20}.
  \item In machine learning outside neural networks, \emph{Bayesian
      networks} \cite{PearlRussell03}, \emph{dynamic Bayesian
      networks} \cite{Murphy02a}, and most generally \emph{graphical
      models} \cite{Jordan04} establish a unifying formal framework
    for representing and learning high-dimensional probability
    distributions and calculating inferences on
    them. \citeA{Pecevskietal11} describe how statistical inferences in
    graphical models can be realized in stochastic spiking neural
    networks. While one standardly considers only inferences of
    conditional and marginal distributions, novel methods for
    identifying \emph{causal} interactions between observables are
    claimed to have the potential to revolutionize machine learning
    because these methods may dramatically reduce the amount of
    training data needed to achieve a desired functionality
    \cite{Schoelkopfetal21}.
  
  \end{itemize}

\item \emph{Connectionism} refers to a hybrid symbolic-dynamical way
  of thinking about neural networks, which has been understood
  differently at different times. Here we point out
  \emph{spreading activation} models of cognitive neural
  processing. These models are structured as graphs whose nodes are
  labeled with concept or operator names, with activating or
  inhibiting links between them. These formalisms are a crossover between
  symbolic AI (because of the symbolic node labels) and the parallel
  distributed neural processing paradigm (due to the continuous-time
  activation-based interaction). A deeply worked-out, exemplary family
  of connectionist models are the SHRUTI models of linguistic
  processing, which are \emph{``neurally motivated models of relational
  knowledge representation and rapid inference using temporal
  synchrony''} \cite{Shastri99a}.
  
\item \emph{Program engineering for spiking neurochips.}
  \citeA{Zhangetal20} present a method for engineering brain-inspired
  computing systems, programming them in a high-level formal design
  language, which is compiled down through an intermediate formalism
  to a machine interface level, which then can be mapped to the
  current most performant (digital) neuromorphic microprocessors. This
  approach is motivated by practical system engineering goals and in
  many ways follows the role model of AC compilation hierarchies.  It
  is however limited to exactly the three specific modeling levels
  specified in this work, with different principles used for the
  respective encodings, and at the bottom end exclusively targets
  digitally programmable spiking neurochips.

\item The \emph{Realtime Control System} \cite{Albus93} of James Albus
  is a design scheme for control architectures of autonomous robotic
  systems, from the sensor-motor interface level to high-level
  knowledge-based planning and decision making. Like other models of
  modular cognitive architectures \cite{Samsonovich10}, it is taken for
  granted that they are simulated on digital computers.

\item In theoretical physics, \emph{[pan-]computationalism} loosely
  refers to a variety of modeling approaches where formal concepts
  from symbolic computing or the Shannon concept of information are
  invoked to describe and explain the physical world (introduction:
  \citeA{Zenil13a}; examples:
  \citeA{vonWeizsaecker85,Lloyd13,Wolfram20}). A notably popular
  special format are \emph{cellular automata} models, which capture
  self-organized pattern formation in physical substrates
  \cite{Zuse82, Wolfram02, Fredkin13}.

\item \emph{Physarum computing} \cite{Adamatzky18} is a popular
  subject of experimental and theoretical studies in the
  unconventional computing arena. Physarum is a genus of slime molds,
  whose amazing life cycle comprises stadiums of single-cell
  amoeba-like organisms, merged macroscopic multi-nucleus megacells,
  and large multicell-bodies shaped as mycel webs or 
  'mushrooms'. Their individual and collective information-processing
  capabilities are investigated under biological aspects like
  orientation and navigation, but also under pure computational
  aspects --- physarum colonies have been grown into Boolean circuits!

\item \emph{Stochastic search for optimization tasks.} Task-solving
  'computing' is variously framed as solving optimization tasks. When
  the cost landscapes are non-convex, complex and non-differentiable,
  stochastic search methods are the only known way to find good task
  solutions. A spectrum of ensemble-based parallel search methods has
  been proposed in various communities, for instance simulated
  annealing \cite{Kirkpatricketal83}, (classical) DNA computing
  \cite{vanNoortetal02}, or ant colony algorithms
  \cite{DorigoGambardella97}.

\item \emph{Self-assembling DNA macromolecules} \cite{Doty12} is a
  rather recent development in DNA computing. A mix of DNA-based
  nanomolecular complexes, each of which has a specific geometrical
  shape, is put into a reactor, in which these DNA 'tiles' bind into
  growing complexes whose regular shapes encode and process symbolic
  information, akin to what happens in cellular automata or even a
  Turing machine. This model of information processing combines
  aspects of algorithmic processing, stochastic search, and
  self-organized pattern formation. This line of research is mostly
  carried out in computer simulations and abstract mathematical
  characterizations, with limited experimental demonstrations so far.

\item In \emph{hyperdimensional computing} \cite{Kanerva09},
  conceptually interpretable information items are represented by
  (long) random bitstrings, which can be transformed and combined by
  operations that correspond to logical or algebraic operations, and
  support the creation of hierarchically nested information
  structures. Hyperdimensional computing can be regarded as a version
  of \emph{stochastic computing}, which goes
  back to \citeA{vonNeumann56}.

\item \emph{Neural field theory} \cite{LinsSchoener14} formalizes
  neural dynamics in terms of spatiotemporal pattern formation on
 neural sheets. Moving neural 'solitons' can
  represent the activation of concepts. The theory can give an
  integrative account of information
  processing of interacting top-down and bottom-up processing
  pathways across several layers in the neural hierarchy, and has been
  linked to the non-neural, generic machine learning architecture of
  \emph{map seeking circuits} proposed by David Arathorn
  \cite{GedeonArathorn07} --- one of the very few formal models of
  bidirectional (top-down \& bottom-up) information processing
  systems.

\item \emph{Self-assembling DNA macromolecules} \cite{Doty12} is a more
  recent development in DNA computing. A mix of DNA-based
  nanomolecular complexes, each of which has a specific geometrical
  shape, is put into a reactor, in which these DNA 'tiles' bind into
  growing complexes whose regular shapes encode and process symbolic
  information, akin to what happens in cellular automata or even a
  Turing machine. This kind of information processing combines aspects
  of algorithmic processing, stochastic search, and self-organized
  pattern formation.This line of research is mostly carried
  out in computer simulations and abstract mathematical
  characterizations, with limited experimental demonstrations so far. 

\item \emph{Physarum computing} \cite{Adamatzky18} is a popular
  subject of experimental and theoretical studies in the
  unconcenventional computing arena. Physarum is a genus of slime
  molds whose amazing life cycle comprises stadiums of single-cell
  amoeba-like organisms, merged macroscopic multi-nucleus megacells,
  and large multicell-bodies shaped as mycel webs or fructuation
  'mushrooms'. Their individual and collective information-processing
  capabilities are investigated under biological aspects like
  orientation and navigation, but also under pure computational
  aspects --- physarum organisms have been grown into Boolean
  circuits!
  
\item Under the label of \emph{neural-symbolic integration} or
  \emph{neural-symbolic computing}, an interdiscipinary community with
  roots in formal logic, AI, cognitive science, and artificial neural
  networks has found together in a shared effort of interpreting
  neural dynamics in terms of logical information processing, and of
  designing hybrid neural-logical algorithms for learning and
  inference \cite{Besoldetal17}. A broad range of formal models are
  proposed and discussed in this setting. This community is quite
  productive with
  \href{https://www.city-data-science-institute.com/nesy}{founding a
    scientific association, book publications and a yearly workshop
    series}.

\end{itemize}

This list is certainly incomplete. Still, it illustrates the already
existing wealth of formally worked-out perspectives to interpret
natural or engineered systems as 'computing', 'information
processing', or 'cognitive' in some way or other that differs from the
standard model of digital/symbolic/algorithmic computing. Obviously,
none of these approaches has yet established itself as a commonly
agreed unifying theory framework for the entire community of
neuromorphic and other unconventional computing investigators. This is
inevitable when one considers the diversity of scientific or
epistemological goals that gave rise to these efforts. 
In order to tie together
the spreading-out threads of neuromorphic and other unconventional
computing researches --- and thus laying the foundations for a
scientific engineering discipline --- a unifying theory of 'computing'
arising from whatever physics can offer is needed. All of the
currently available formal models of (wide-sense) computing systems
are missing one or more of the following necessary conditions that
such a unifying \emph{general formal theory of physical computing
  systems} (GFT) theory must satisfy:

\begin{enumerate}
\item {\bf Phenomenal openness.} A GFT must provide formal tools to
  express 'computational' functionalities that emerge from a wide
  variety of physical phenomena. It is not enough for a GFT to
  address, exclusively, either multistable switching, or the
  stochastic dynamics of statistical ensemble systems, or
  self-organized pattern formation.
  
\item {\bf Interpretability.} A GFT must include ways to formally
  characterize the use-cases and tasks that can be served by a given
  physical computing system. It is not enough for a GFT to capture the
  'mechanics' of computing systems --- it must also give an account of
  what the mechanical processes 'mean' - their task-related
  semantics.
  
\item {\bf Scalability.} The range of possible tasks which can be
  addressed through GFT models must be very wide. It is not enough for
  a GFT to cover only, for example, supervised learning or stochastic
  search tasks. Furthermore, the complexity of achievable tasks must
  be arbitrarily scalable.
  
\item {\bf Model abstraction.} A GFT must provide rigorous methods for
  model abstraction. For practical system engineering it is
  necessary to describe the workings of a computing system at
  different levels of granularity --- fine-grained and
  hardware-oriented at low levels of abstraction, coarse-grained and
  task-oriented at higher abstraction levels.
\end{enumerate}

The textbook theory of symbolic/digital computing excels with regards
to points 2--4 but fails at 1. As far as we can see, the only approach
to 'computing' which to some extent fulfils condition 1 is reservoir
computing, which explains its popularity in computational materials
research --- but reservoir computing has nothing to offer with regards
to 2--4. Some other approaches from our listings can accommodate
some but not all of our four requirements, but many fall short
in all regards 1--4.

We have left out one interesting objective, the most classical of all:
the demand that computational operations must be \emph{effectively}
realizable.  In the mathematical tradition of thinking about
'computing', qualifying a numerical function as 'computable' was
tantamount to requiring that there be an 'effective' method to
calculate its output values. The meaning of this term is intuitively
clear to mathematicians, boiling down to the vision of a mathematician
who uses paper and pencil to write down a sequence of formulas
according to mathematically correct transformation rules, until the
last formula shows the result. It remained however for Turing to cast
these intuitions into a precise mathematical model of 'effectiveness',
namely the Turing machine. The Turing machine definition of
'effectively computable' functions is today widely regarded as the
ultimate answer --- any function that can be computed by \emph{some}
machine can also be computed by a Turing machine.  Discussions in the
\emph{hypercomputing} community \cite{Copeland02, Ord06} have not yet
led to convincing propositions of functions, which could be computed
by some sort of physical machine but not by a Turing machine.
However, all of these discussions are concerned with a specific class
of input-output transformations, namely mathematical functions
$f: \mathbb{Z} \to \mathbb{Z}$ from integers to integers (or,
equivalently, from finite symbolic data structures to finite symbolic
data structures). These functions may be only partially defined (no
result value defined for some input arguments), but for arguments
where they are defined, the function output value is unique, precise,
and finitely specifiable. This excludes from the discussion other
sorts of input-output transformations that could be nondeterministic,
stochastic, have only partially defined results, or have results that
cannot be specified in a finite description, or are context- or
time-dependent. While such input-output relations are not mathematical
functions, they may well be relevant for practical 'computing'
systems. They are also characteristic of the input-output
functionalities of biological brains, the archetypical 'computing'
system for the neuromorphic community.  In a lecture delivered in the
year 1948, John von Neumann himself, focusing on the stochasticity and
error tolerance of biological neural processing, found the Turing
model of 'computing' inadequate and concluded that \emph{``we are very
  far from possessing a theory of automata which deserves that name,
  ...''}.  A full understanding of 'computing' would require a new,
\emph{``detailed, highly mathematical ... theory of automata and of
  information''}, and such a theory would likely contain elements of
\emph{``analytical''} (continuum, calculus) mathematics
\cite{vonNeumann48}, which are alien to the Turing paradigm. Besides
brains, other natural systems from ant colonies to slime molds are
today viewed by many as 'computing' systems --- indeed, in the
unconventional computing communities these are variously taken as role
models. In summary, the traditional understanding of what makes
computational procedures 'effective' is tied to discussing 'computing'
only in the sense of evaluating mathematical functions on the
integers. This shuts the eyes on many aspects of information
processing in natural systems --- and coming to terms with these
aspects we consider as highly relevant for for neuromorphic and
unconventional theory-building.

In this article we propose yet another approach toward a general
formal theory of computing systems. Our strategy is to let all
considerations start from this first goal of physical openness, and
gratefully accomodate whatever guidance we can get from classical
computer science theory with regards to the other three objectives. In
the end, we hope to come out with a schema for a GFT that comprises
the classical symbolic-algorithmic theory one as a special case, which
would be characterized by a confinement of physical phenomena to
binary state switching dynamics.

Our role model for a practically relevant theory of computing systems
is --- inevitably --- the textbook theory body of digital/smbolic
computing. This is not a single philosopher's stone but a cosmos of
interrelated subtheories. While the classical theory of symbolic
computing is often referred to as the theory of Turing computability,
this is a metonymic usage of terminology. Besides the Turing machine
(or any other equivalent model of algorithms), textbooks of
theoretical computer science minimally comprise subtheories of formal
languages and automata, computability and complexity, and formal
logic. This classical compendium of subtheories leads to the amazing
interpretability, scalability and abstractability of digital/symbolic
computing models. Likewise, for a GFT we envision not a singular, compact
formalization of a physical computing system, but a microcosm of
interrelated formal theories and models. While this makes the
task bigger and more complex, we believe that a differentiated
compendium of subtheories is needed to meet the expansive demands of
our goal quartet.

This article is structured as
follows.

\begin{itemize}
\item We begin by highlighting the richness of physical materials and
  phenomena that could possibly be exploited for engineering computing
  systems (Section \ref{secGifts}, \emph{The gifts of Mother Physics}).
\item We argue that the classical body of digital/symbolic models of
  computing systems, which have been developed in the succession of
  Turing, are ill-suited as a basis for developing theories of
  physical computing. Turing's concept of 'computing' roots in
  abstract logical reasoning mechanics, and the realization of logical
  reasoning mechanics in physical machines restricts the exploitable
  physics to finite-state switching dynamics --- the 1-0 bit values
  are reflections of the logical True-False values. We argue that
  a theory of physical computing should be grown bottom-up from the
  physical phenomena that one recruits for computing (Section
  \ref{secNotTuring}, \emph{Alan Turing was a mathematician, not a
    physicist}).

\item We propose a universal organization schema for a stack of formal
  modeling levels which we think are needed for any general theory of
  physical computing. These modeling levels span from physical models
  of physical hardware systems through interconnected procedural
  computing models to declarative task models (Section
  \ref{secStrucPhysTheory}, \emph{The structure of theory systems for
    physical computing systems}).
\item We take a closer look at this universal schema and work out two
  more specific and partial instantiations, one for the classical
  symbolic/digital view and one for the cybernetic view that we put
  forward as better suited for a bottom-up capture of arbitrary
  physical phenomena (Section \ref{secIntroCybernetic},
  \emph{Algorithmic and cybernetic theory hierarchies}).
\item We identify a few fundamental questions that need to be answered
  by any general theory of physical computing systems (Section
  \ref{secChallenges}, \emph{Big challenges ahead}).
\item We propose a particular strategy for developing a GFT, which we
  call 'fluent computing'. The core idea is to reverse the top-down
  perspective of Turing computability --- which starts from Turing
  computability as logical/symbolic reasoning and breaks this down to
  bistable physical switching devices --- and instead start from
  physical observables and assemble a computational modeling hierarchy
  bottom-up from them (Section \ref{secFluentComp}, \emph{Fluent
    computing}).
\item We argue that the classical textbook theory of digital computing
  can be seen as a special instantiation of our proposed fluent
  computing framework (Section \ref{secAlgAsFluent}, \emph{Algorithmic
    theories seen as fluent theories}).
\item We conclude with a brief summary and highlight the benefits that
  a worked-out GFT would bring for founding a systematic engineering
  discipline of computing with general physical systems (Section
  \ref{secConclusion}).
\end{itemize}

\section{The gifts of Mother Physics} \label{secGifts}

A key objective in physical computing is to understand how, given a
novel sort of hardware system made from 'intelligent matter'
\cite{Kasparetal21}, one can ``exploit the physics of its material
directly for realizing its operations'' \cite{Zauner05}. One may hope
that, compared to electronic digital computers, exploiting
unconventional physical effects can enable important savings in energy
consumption. A salient example is the realization of synaptic weights
in neuromorphic microchips through memristive devices
\cite{YangStrukovStewart13}. In digital simulations of neural
networks, updating the effect of a synaptic weight on a neuron
activation needs hundreds of transistor switching events. In contrast,
when a neural network is realized in a physical memristive crossbar
array \cite{Lietal19}, one obtains an equivalent functionality through
a single pulse of a small current that passes the corresponding
memristive synapse element.

This principle of direct physical mirroring is not limited to updating
single numerical quantities, and the potential benefits are not
restricted to energy savings. Other potential advantages include
higher processing speeds (as in optical computing) or higher data
throughput rates due to physical parallelism (as in physical reservoir
computing) or damage robustness (as in brains). Complex
information-carrying formal structures and computational
operations on them --- like inferences on hierarchically defined
concepts, graph transformations, finding minima in cost landscapes,
etc.\ --- can be mirrored in spatiotemporal physical phenomena
in many ways. In turn, these physical phenomena can be modeled by a
range of mathematical constructs. Often these constructs are quite
generic and can be found in almost every sufficiently complex physical
or neural system --- for instance oscillations, chaos and other
attractor-like phenomena; hysteresis; many sorts of bifurcations and
input-induced transits between basins of attraction; spatiotemporal
pattern formation; intrinsic noise; phase transitions. The pertinent
literature for each of them is so extensive and diverse that it defies
a systematic survey. Other mathematical constructs, which have been
discussed as carriers for information processing operations, are more specific,
for instance heteroclinic channels and attractor
relics \cite{Rabinovich08,Gros09}, self-organized
criticality \cite{Chialvo10,Stiegetal12,BeggsTimme12} or solitons and
waves \cite{LinsSchoener14,Grollieretal20}.  Fabricating materials with
atomic precision is today routinely done in tens of labs worldwide,
exploring optical, mechanical, magnetic, spintronic or quantum effects
and their combinations, for instance in nanowire
networks \cite{Stiegetal12,KuncicNakayama21} or skyrmion-based
reservoir computing \cite{Leeetal23}. Physical materials and devices
offer a virtually limitless reservoir of physical phenomena for
building unconventional computing machines.

An illustrative example comes from our own work.  B.N.\ investigates
ferroelectric and ferromagnetic effects in novel computational
materials. These materials display an ordered phase, which is
responsible for long-term bi-stability, and a disordered phase, in
which these properties vanish. The transition between these two phases
takes place because the structure-forming electrical or magnetic
forces between neighbouring atoms compete with the entropy of the
system. Ordering across macroscopic distances gives rise to strong
nonlinear responses to external stimuli. The complexity and
sensitivity to external stimuli is maximized at phase
transitions \cite{Langton90,Kinouchi2006}. Some of the novel materials
that we synthesize combine multiple types of interactions (magnetic,
electrical, mechanical, chemical) and thus display complex phase
diagrams with multiple available phases. Using state-of-the-art
thin-film deposition techniques, we can make materials that persist
permanently at the edge between two phases, or close enough to a phase
transition \cite{Everhardt2020}, such that they can be brought from one
phase to the other with a small external
stimulus \cite{Everhardt2019}. Such materials at the verge of stability
present rich energy landscapes with multiple metastable states that
can be switched with low energy expenditure.  In these materials,
self-assembly of ordered domains and their domain walls takes place as
the ordered phase grows from the disordered phase.  The evolution of
these topological defects often results in hierarchical structures,
which in one study we observed as periodicity doubling cascades, a
signature of spatial chaos \cite{Everhardt2019}.

\begin{figure}[htb]
  \center
  \includegraphics[width=6cm]{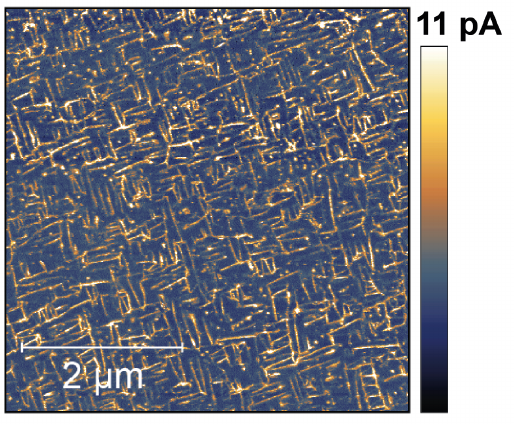}
  \caption{Conduction map (obtained using conducting force microscopy)
    of a ferroelectric BiFeO$_3$ thin layer, showing a structure of
    insulating ferroelectric domains (blue) and conducting domain
    walls (yellow). The image was obtained using conducting force
    microscopy at room temperature, taken from
    \protect\citeA{Riecketal22}}
  \label{figDomainWalls}
\end{figure}

Figure \ref{figDomainWalls} shows
a network of conducting domain walls in a ferroelectric BiFeO$_3$ thin
layer (thickness approx. 50 nm). Electrical currents can be
transmitted across the domain wall network (yellow), while the regions
in between the walls are insulating ferroelectric domains that can be
switched into non-volatile memory states (blue). 

The properties of these materials --- fine-grained conductivity
pathways, local multi-stability with resistive properties that are
switchable with minimal energy, many timescales, hierarchical
topological structuring --- hold many promises for computation in
materials directly.  Spatiotemporal processes in a regime 'at the edge of
criticality' near a phase transition have been described as an
enabling condition for complex information processing that use such
dynamics \cite{Chialvo10,BeggsTimme12}. More concretely, structures
like these BiFeO$_3$ thin layers and other substrates, which support
very large numbers of switchable and energetically interacting memory
states, might become used for example to encode and dynamically switch
the large random bit vectors which are the main representational
elements in hyperdimensional computing \cite{Kanerva09} or in Hopfield
networks, Boltzmann machines and other 'Ising machines'
\cite{Hopfield82,Ackleyetal85, MarulloAgliari20,
  Kiralyetal21,Zhangetal22}. 

Unconventional materials and phenomena may have fascinating potentials
--- but whether or how these can be turned into practically relevant
computing machineries and applications remains to be seen.

\section{Alan Turing was a mathematician, not a
  physicist}\label{secNotTuring}

One may think that we already have a shared, general formal concept of
'computing', namely what a Turing machine can do. The Turing model of
'computing' stands out among all other existing models. Even
philosophers, when they try to come to terms with the essence of
'computing', invariably orient their argument toward Turing
computability \cite{Harnad94, Piccinini07}. Its mathematical theory
has been worked out far wider than any other theory of 'computing'; it
is elegant and transparent; it is the most technologically productive
one; and it connects 'computing' to formal logic, which in turn yields
a semantic theory of computing processes, which allows us to formally
specify what a 'computation' \emph{means} with regards to solving
real-world tasks. One of the consequences of the intimate connection
of Turing computing with logic is that at the lowest implementation
level of digital computing systems one finds the most elementary level
of logical formalisms, namely Boolean logic. Digital 'computing'
models and hardware systems are built on the basis of Boolean logic gates.
Almost universally across unconventional and neural modeling
approaches, researchers describe ways how Boolean gates can be
realized within the respective non-digital formalism. The very first
comprehensive formal abstraction of brains casts it as a Boolean
network \cite{McCullochPitts43}. For decades, learning the XOR
function was a touchstone for artificial neural networks
\cite{MinskyPapert69}, and re-creating Boolean functions still is a
standard proof of power for unconventional computing
\cite{Adamatzky15, Boseetal15}. Efforts to implement Boolean logic in
optics have been a main objective in optical computing
\cite{BrennerHuangStreibl86}. Emulating Turing machines in neural
networks was an acclaimed achievement in the deep learning world
\cite{Gravesetal16}. And more often than not, formal analyses of
'computing' done  in the unconventional computing community relate back to
concepts from the symbolic/digital Turing theory, for instance when
discussing the potential powers of unconventional computing machines
in terms of classical complexity theory \cite{Blakey17}. Finally, and most notably, every physical system, for which physicists can offer an formal model of its states and dynamics, can be simulated on digital computers with arbitrary precision.

Given these powers and beauties of the Turing paradigm, and its
undeniable role as an anchor reference for all discussions of
'computing', why should one wish to develop a separate theory for
'computing' based on general physical phenomena at all? What could such a
theory give us that Turing theory cannot?

We begin with the historical roots of the Turing machine
concept. With this formal concept, Turing set the capstone on two
millenia of inquiry which started from Aristotle's syllogistic logic
and continued through an uninterrupted lineage of scholars like
Leibniz, Boole, Frege, Hilbert and early 20th century logicians. The
original question asked by Aristotle --- what makes rhetoric
argumentation irrefutable? --- ultimately condensed in the
\emph{Entscheidungsproblem} of formal logic: is there a effective
logical/mathematical method to decide (that is, find a mathematical
proof) for every mathematical conjecture whether it is true or false?
While all the pre-Turing work in philosophy, logic and mathematics had
finalized the formal definitions of what are conjectures, formal
truth, and proofs, it remained for Turing to give a formal definition
of what is an 'effective' method for finding proofs --- for 'computing'
proofs --- namely, that 'effectively' carrying out a computation is
equivalent to running a Turing machine.

Turing's background concepts did not grow from intuitions about
physical computing systems. He distilled his machine concept as part of
a solution to a specific, deep problem of formal logic.  In his famous
article \emph{On computable numbers, with an application to the
  Entscheidungsproblem} \cite{Turing36} --- which in retrospect
laid the foundation for today's computer science --- Turing
conceived of his formal machine model as an abstraction of mathematical
thinking. He concretely and explicitly describes how his Turing machine
abstracts from a human (male) mathematician who, equipped with paper
and pencil, does his mathematical thinking job. The
Turing machine consists of a tape on which a stepwise moving cursor
may read and write symbols, with all these actions being determined by
a finite-state switching control unit. The tape 
models the sheet of paper used by the mathematician, the machine's
read/write cursor models his eyes and hands, and the finite-state
control unit models his thinking acts.  Citing from that famous article:

\emph{``Computing is normally done by writing certain symbols on
  paper. [...] I shall also suppose that the number of symbols which
  may be printed is finite. If we were to allow an infinity of
  symbols, then there would be symbols differing to an arbitrarily
  small extent j. The effect of this restriction of the number of
  symbols is not very serious. It is always possible to use sequences
  of symbols in the place of single symbols. [...] The behaviour of
  the computer at any moment is determined by the symbols which he is
  observing, and his 'state of mind' at that moment. [...] We will
  also suppose that the number of states of mind which need be taken
  into account is finite. The reasons for this are of the same
  character as those which restrict the number of symbols. If we
  admitted an infinity of states of mind, some of them will be
  'arbitrarily close' and will be confused.''}

Turing deliberately decided for a discrete, even finitary mathematical
format for his model of 'computing'. We may suspect that besides the
outward reason that he gives in his article for this decision (the
practical inseparability of infinitely many symbols or processing
states) he also had in mind his ultimate objective of formalizing the
processing of logical formulae, which are combined from a finite set
of symbols and processed with a finite set of logical derivation
rules. This decision to opt for finite symbol and state sets is
supremely productive, as the history of DC shows. But this decision is
also very restrictive. It bars the way to an immediate mathematical
grasp on all 'computing' processes that are continuous in state and
time, and/or are inherently stochastic, and/or are spatially organized
in physical 3D substrates --- thus excluding almost all natural
'computing' systems.

Importantly, Turing speaks of \emph{``states of mind''} when he refers
to the switching states of the control unit. He does not speak of
physiological brain states. The Turing machine models reasoning
processes in the abstract sphere of mathematical logic, not in neural
physiology. Students of symbolic/digital computer science must do
coursework in formal logic, not physiology; and their textbooks speak
a lot about logical inference steps, but never mention seconds (we
will use the acronym CS from now on to refer to the classical textbook
body of symbolic/discrete computer science).

A brief philosophical aside: Different mathematicians can have
different views on the ontological domains in which mathematical
objects or concepts
exist. A platonist mathematician will believe that
  mathematical objects exist in a reality that comes before the
  physical reality; a constructivist will believe that these objects
  exist to the degree that mathematicians can effectively define them
  together with the operations that can work on them; an intuitionist
  will think of mathematical objects as existing in the minds of
  mathematicians; and there are more such views \cite{Spalt81}. These
  differences are of no concern for viewing the Turing machine as a
  model of a thinking mathematician. One might say that Turing himself
  was an intuitionist because his considerations center on ``states of
  mind'', but we prefer to leave the classification of
  Turing's philosophical stance to expert philosophers.

Turing's commitment to the `thinkability' of only discrete symbols
was rooted in two millenia of Aristotelian philosophizing about
logical rational reasoning. In this tradition, the existence of
discrete concepts is a primary given (as opposed to the Heraclitean
tradition and Christian mysticism, schools of thought that
historically evolved in parallel). This Aristotelian
heritage has also left its mark on linguistics and neuroscience, where
it leads to the important and difficult question how analog neural
systems can ``think'' discrete concepts and symbols. This is often
mathematically modeled by attractor-like phenomena in nonlinear neural
dynamics \cite{Durstewitzetal00, PascanuJaeger10, FusiWang16,
  Jaeger21a}.

When one understands the
Turing machine as a general model of rational human reasoning, it
becomes clear why digital computers can be simulation-universal:
everything that physicists can think about with formal precision can
be simulated on digital computers, because these machines can simulate
a physicist's formal reasoning.

The royal guide for shaping intuitions about the physical basis of
'computing' are biological brains, especially human brains. We will
now take a closer look at what aspects of neural information
processing are not captured by Turing computability.

The first thing to note is that the word 'digital' is used outside CS
sometimes in a quite generic way, pointing to anything that appears as
'binary', 'yes-no', 'discrete-event-like', etc. Used in this wide
sense, neural spikes and vesicle releases are frequently said to be
`digital'. 

However, within CS, the word 'digital' has a clearly defined meaning
which is much more specific than that wide-sense use. Neural spikes or
vesicle releases do not qualify as 'digital' in this specific CS sense
of the word. We will explain this in some detail, because it is an
important issue.

Digital computers are designed and programmed according to the
principles of algorithmic computing theory. This theory is
paradigmatically represented by the Turing machine model, which
outside CS is the best-known model. CS textbooks however introduce
numerous other formal models of algorithmic computing, e.g. lambda
calculus, general grammars, logic programming, random access machines,
cellular automata. They are all equivalent to Turing machines in the
sense that they characterize the same set of computationally solvable
input-output functions. The common denominator is that they are all
based on finite sets of symbols --- 0 and 1 in the special case of
digital machines. All these equivalent formal models of algorithmic
computing describe how symbols can be hierarchically composed into
complex structures (also called 'expressions', 'words', 'data
structures', 'formulas', 'programs', 'files' etc.\ depending on the
context), and how these expressions can be transformed in discrete
update steps according to a finite set of finitely specified
rules. Specific, likewise finitely specifiable meta-rules for
sequencing such rules are 'algorithms'. When a computer scientist
speaks of 'digital computing', this is metonymic usage and stands for
the general principles of symbolic/algorithmic computing regardless
whether the fundamental underlying symbol set is the digital symbol
set $\{0, 1\}$ or something else, for instance the 128-element ASCII
alphabet that is used in programming languages or the digits 0--9 used in Babbage's historical arithmetic calculator machines. 

The root concept of algorithmic/symbolic/digital computing is the
concept of a symbol. There exists no mathematical definition of what a
'symbol' is - mathematicians use this concept as a primary given and
can rely on a robust, shared, intuitive understanding. Tacit
assumptions about 'symbols' are that they are individually
recognizable with perfect certainty; that they can be ``written''; and
that when they have been written they persist for arbitrary long times
until they might be overwritten; and that they can be composed
into words, which by definition are finite sequences of symbols. Any other data structure, like trees or tables or entire
computer programs, can be 1-1 encoded into and decoded from words,
thus the textbook theory of CS typically is based entirely on words. 
Symbolic 'computing' processes are chains of discrete transformations
of words, from an input word to an output word. It is important that a word remains
immutable for indefinite timespans after it was 'written', until it
gets overwritten by some transformation operation. The immense power
and mathematical as well as philosophical beauty and depth of the
textbook theory of CS lies in the fact that these transformation rules
can be characterized and semantically interpreted by formal logic
systems, typically first-order logic, which every CS student has to
learn \cite{Schoening08}. This allows to formally specify input words
as semantically interpretable encodings of 'tasks' that are specified
in a formal logic, which in turn allows CS analysts to formally prove
that an algorithm does indeed precisely solve the intended task ---
this is the branch of CS called 'program verification'. There is a
stringent connection of this logic interpretation of symbolic
computing processes with the fact that digital computers (here we mean
indeed the 0-1-based digital machines that surround us) operate at the
lowest level with logic gates --- these circuits implement the most
elementary logically interpretable word update rules from which all
others can be assembled.

The operations of a brain can be cast as algorithmic/symbolic/digital
computing in two ways. First, one can view the brain's operation as
algorithmic. In this view, the tasks that a brain can solve are the
tasks that digital computers (or any other equivalent algorithmic
machine) can solve. This is the view taken in the classical work of
\citeA{McCullochPitts43} who model a brain as a network of logic
gates. The McCulloch-Pitts neuron model allows for two activation
states 1 and 0 (or active versus inactive), which are identified
with the True / False values of Boolean logic variables. A
McCulloch-Pitts neuron model is digital in the CS sense. Once switched
to, say, a 0 value, it immutably preserves this value until its state
is switched by new input. This has nothing to do with spikes
--- the landmark paper of McCulloch and Pitts does not mention spikes
at all. If one wishes to map the abstract McCulloch-Pitts neuron to
biological neurons --- a very far stretch --- their 0/1 values would
become neuron activation values, not spike events. A McCulloch-Pitts
brain is a logical reasoning machine.

The second way how a brain can be digitally modeled 
is to use a digital computer to numerically simulate the
analog, continuous real-time, spatiotemporally organized, stochastic
processes observed through physiological measurements. Then, brains
are modeled \emph{by}, not \emph{as}, a digital system. This is in
principle possible to any desired degree of approximation. Such
simulation models do not give an account of the brain's operations as
symbolic-logical inferences. These simulations are just another case
of using modern computers to simulate some physical system. They are
good for experimentation-by-simulation, but they do not model (let
alone explain) the information processing functionality of the
simulated brain. Unfortunately (and importantly), the further such
simulations are refined for greater accuracy, the slower they become.
A spectacular, prize-winning recent example for the slowness of
high-accuracy simulations of physical systems is the simulation of
surface protein reconfigurations in the SARS-Cov2 virus, which kept
the worldwide second-largest supercomputing cluster busy for days to
simulate nanoseconds of molecular dynamics \cite{Dommeretal21}. The
slowness of realistic simulations of physical dynamics quickly becomes
crippling also in much more modest scenarios. In own work of co-author
HJ, we tried to understand and predict the dynamics of the DYNAP-SE
analog spiking neurochip \cite{Moradietal18} by simulation on a
high-end PC, using the BRIAN software. We gave up because the
simulations were orders of magnitude slower than the simulated
real-time despite significant modeling simplifications, and we
reverted to physical experimentation and measurements of the actual
chip \cite{Heetal19}.

The McCulloch-Pitts approach (and later, the view held by the
proponents of the physical symbol systems paradigm) consider the
information-processing functionality of brains \emph{as}
digital/symbolic/algorithmic systems. This approach reflects (and is
limited to) a specific understanding of 'information', which at its
core is the assignment of truth values to logic expressions.  In
contrast, numerical simulations model the procedural mechanics of
brains \emph{by} digital/symbolic/algorithmic tools. Such simulations
do not demonstrate or claim that the information-processing
functionality of brains is the same as algorithmic computing; they do
not give an account at all of what the 'information' is that is being
processed --- this interpretation is external to the numerical mechanics
of the simulation engine and must be supplied by the human researcher.

If one wishes to understand in what sense brains 'process information'
(or 'compute'), one needs a conceptualization of 'information
processing' for starters. One can assume (or believe or argue) that
the symbolic/algorithmic/logic-based conceptualization is the right
one --- then the McCulloch-Pitts approach is the way to go, in its
original version or in one of the later, more sophisticated variants
of symbolic, logic-based AI. The challenge then is to find correlates
of writable, immutable symbols in neural dynamics, and neural
mechanisms that can be regarded as realizing discrete logical update
rules. If one then wants to build brainlike machines, digital
computers already do it natively and naturally, and there is no
incentive to invest further thought in unconventional computing
systems. However, we believe that the algorithmic interpretation of
'information processing' is only partially relevant for understanding
brains, namely for modeling the high-end rational reasoning
functionalities, which only few animal species have developed and
which even in humans is very far off from the perfect powers of formal
algorithmic computing.

The mathematical tools for modeling symbolic information processing
had originally been developed by linguist Noam Chomsky, who created
them in order to analyse the structure of natural language and the
mental operations executed by the human brain.  In the high times of
classical symbolic AI there was a heated philosophical debate about
whether viewing brains as logical reasoning machines is an appropriate
or even the only appropriate way of understanding neural information
processing --- or whether this is an undue simplification which
prevents us from understanding biological brains and from engineering
computing systems that truly deserve to be called `intelligent'. The
former view found its most pointed expression in the 'physical symbol
systems hypothesis’ stated and defended with authority by
\cite{NewellSimon76}.  Proponents of the latter position argue that
the ultimate sources of biological intelligence must be sought in the
apparent stochasticity and analog gradedness in human perception,
reasoning, speech and action; that one cannot understand cognition as
disembodied rational reasoning but that instead it realizes itself in
the embodied situatedness of intelligent agents in continual physical
interaction with their environments; and that self-organizing
continual adaptation and learning are the key to cognition. In these
views, complex symbol structures are an epiphenomenon and, for that
matter, only very imperfectly realized in biological
cognition. Arguments of this sort have been put forward in the
philosophy of mind \cite{Clark13}, cognitive and evolutionary
linguistics \cite{BickertonSzathmary09, Soleetal10}, cognitive science
\cite{Bartlett32,Lakoff87,GelderPort95, Hofstadter95}, artificial
neural networks \cite{RumelhartMcClelland86}, robotics
\cite{Brooks91a} or `New AI' \cite{PfeiferScheier98}. We remark that
research in neuromorphic and unconventional computing mostly unfolds
in the spirit of this second perspective, although most
researchers in these fields today do not engage anymore in
those epistemological debates. The Turing machine certainly is a bold
abstraction of a particular aspect of a human brain's operation.

Most of the time most parts of our brain are not busy with logical
reasoning. In our lives, most of the time we do things like walking
from the kitchen table to the refrigerator. Yet, the kitchen-walker's
brain is thoroughly busy with the continual processing a massive
stream of sensor signals, smoothly transforming that input deluge into
finely tuned, uninterrupted command signals to hundreds of muscles. We
like to call this processing of sensorimotor flows of information the
'cybernetic' mode of computing. For the largest part of biological
history, evolution has been optimizing brains for cybernetic
processing --- for \emph{``prerational intelligence''}
\cite{CruseDeanRitter13}. Only very late, some animals' brains
acquired the additional ability to detach themselves from the
immersive sensorimotor flow and generate logico-symbolic reasoning
chains. Several schools of thinking in philosophy, cognitive science,
AI and linguistics explain how this ability could develop seamlessly
from the cybernetic mode of neural processing, possibly together with
the emergence of language
\cite{Bradie86,Greenfield91,Drescher91,PfeiferScheier98,LakoffNunez00,FedorIttzesSzathmary09}.

\section{The structure of theory systems of physical
  computing systems} \label{secStrucPhysTheory}

In the Introduction we argued that a practically useful formal
framework for computing systems will likely not come in the form of a
single core theory (like the theory of Turing machines), but that
instead it will have to consist of a network of interrelated
subtheories that span the modeling levels from physical hardware to
task specification formalisms (like the theory canon of CS textbooks,
which contains the Turing machine model as one among other
subtheories). In this section we propose a general schema for such a
hierarchy of subtheories. This schema is not itself a theory of
physical computing systems but an organization plan to design
them. Its usefulness lies in making us aware of the multiple facets of
the theory-building problem, and in clarifying how different
subtheories should interconnect. 

To distil such a general schema of subtheory organization, we started
from previous studies by \citeA{Horsmanetal14} who proposed an
abstract model of computing systems, intended to capture all currently
discussed sorts of ’computing’ systems; and from \citeA{Jaeger21a} who
analysed and unified three quite different kinds of modeling
frameworks for information-processing systems --- the classical CS
theory body; probabilistic models based on stochastic sampling; and
approaches based on self-organizing adaptive dynamical
systems. Together these two studies converge to a picture of general
'computing' system models that is drawn around three fundamental
requirements:

\begin{enumerate} 
\item modeling 'computing' systems must include modeling their
  \emph{physical basis};
\item there must be formalizations of \emph{information processing
    mechanisms} which transform input information to output
  information (and there must be models of input and output 'data' in
  the first place);
\item modeling 'computing' systems must include a semantic subtheory
  that allows us to formally specify the real-world tasks served by
  the computing system and that gives an account of the \emph{meaning}
  of computational operations.
\end{enumerate}

Together with the four goals that we mentioned in the Introduction
(phenomenal openness, interpretability, scalability, abstraction),
these three theory-architecture requirements present a set of
constraints that is not easy to accomodate.

\begin{figure}[htbp]
  \center 
\includegraphics[width=10.5cm]{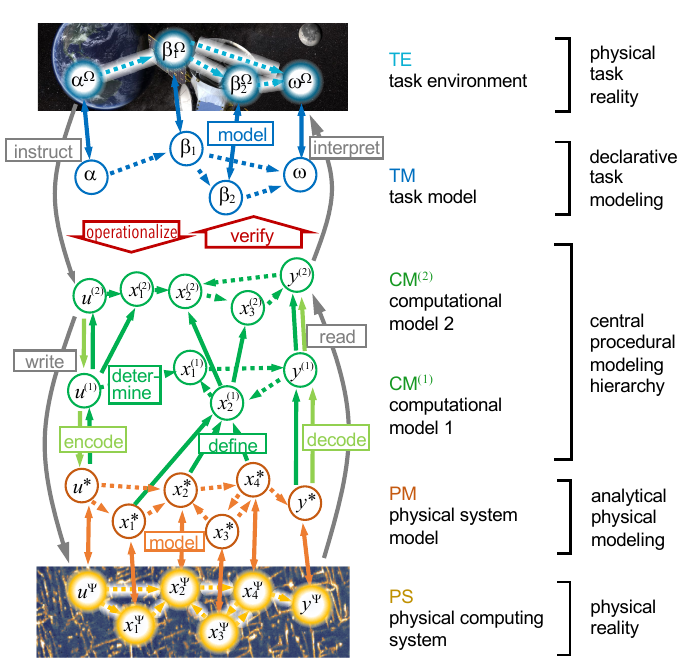}
\caption{An organization schema for a complete modeling hierarchy of
  computing systems. The central element is a hierarchy of
  computational models CM$^{(m)}$, which formalize transformation
  processes from input encodings $u^{(m)}$ through intermediate
  information-carrying states $x_i^{(m)}$ to output encodings
  $y^{(m)}$ (two levels shown). This central procedural modeling
  hierarchy is sandwiched between two sides of physical reality: the
  physical computing system PS and the real-world task environment TE,
  to which it interfaces through an analytical physical
  system model PM and declarative task model TM.  (Top background
  image from \protect\citeA{MITNews19})}
\label{figModelStructure}
\end{figure}

The studies by \citeA{Horsmanetal14} and
\citeA{Jaeger21a} present box-and-arrow diagrams for the
structure of theory systems for computing systems. We merged structural
ideas from both sources and added new elements and detail in order to
make it more concrete and instructive. An overview of our extended
schema is given in Figure \ref{figModelStructure}.

The circled items and arrows in this figure mean different things in
the various layers.  In the physical computing system PS we posit
physical input states $u^\Psi$, intermediate machine states $x_i^\Psi$
and output states $y^\Psi$. Our use of the word `state' needs an
explanation. Physicists speak of the state of a system when they
mean the totality of the system's physical condition at a given moment
in time \cite{Zadeh69}. We call this a global (or total, or system)
state. In contrast, the circled nodes in the PS layer in our diagram
represent partial states --- aspects or components of the global state
that can be spatially localized or otherwise isolated and physically
measured, at least in principle. At different moments in time these
local states can yield different measurement \emph{values}. For
instance, some $x_i^\Psi$ could be a contact point in an electronic
circuit where voltages can be measured; or it could physically extend
to an entire microchip of which the overall temperature is
measured. The yellow broken arrows in PS represent physically causal
influences.  All circles in layers PS, PM, CM$^{(m)}$ are meant to
denote partial states. For simplicity we will just say `state' when
we mean partial states. When we do not want to distinguish between
input/intermediate/output states $u, x, y$, we use the generic symbol
$v$.

The physical system model PM hosts 1-1 formal representatives $v^\ast$
of physical states $v^\Psi$. These $v^\ast$ are abstract, formal items
--- state \emph{variables}. The broken brown arrows in PM stand for
formal models of causal interactions. Often this will be the couplings
between state variables in systems of ordinary differential equations
(ODEs). All variables $v^\ast$ in PM are time-dependant with regard to
the standard model of time $t[\mbox{sec}] \in \mathbb{R}$ that is used
in the natural sciences; thus we could also write $v^\ast(t)$. The
relation between PS and PM is the \emph{scientific modeling} relation
of the natural sciences and engineering: physical states $v^\Psi$ must
in principle be measurable, and the causal interaction models in PM
must lend themselves to generate falsifiable predictions about the
states $v^\Psi$. The model PM is created by hardware engineers or
natural science experts (for instance biologists if PS is a biological
substrate). PM is an \emph{analytical} model (as opposed to a
\emph{blackbox} model), in that it aims at capturing the causal
mechanisms that according to expert insight are active in the material
computing system.

The state variables $v^{(1)}$ in the first computational model
CM$^{(1)}$ are formally \emph{defined} in terms of $v^\ast$ variables
(solid dark green arrows). There may be more or fewer state variables
in CM$^{(1)}$ than in PM. CM$^{(1)}$ is equipped with stochastic or
deterministic, possibly recursive transformation rules which
ultimately \emph{determine} the values of the output variable
$y^{(1)}$ from values of the input variables $u^{(1)}$ via
intermediate variables $x_i^{(1)}$ (broken green arrows). The values
of these variables can be mathematical objects of many sorts: numbers,
vectors, symbols, distributions or other set-theoretic constructs.
Determination pathways may contain cycles. When we say 'determine' we
do not necessarily mean a deterministic transformation relation, but
intend any sort of an effective fixation of output values. Thus, the
broken arrow from $u^{(1)}$ to $x_1^{(1)}$ in Figure
\ref{figModelStructure} may refer to a deterministic function, or a
probabilistic law, or a non-deterministic choice between possible
values of $x_1^{(1)}$.  As a special condition, $u^{(1)}$ must be
defined exclusively from $u^\ast$ and $y^{(1)}$ from $y^\ast$.

In the digital world, CM$^{(1)}$ would be the direct machine
interfacing layer where the machine instructions are resolved to bit
switching operations determine the values of binary state variables
$v^{(1)}$ (this model is used by the microchip engineers but usually
not communicated to customers or programmers).

Higher-up computational models CM$^{(m)}$, if present, are defined
from the respective next lower model CM$^{(m-1)}$ in a similar
way. Figure \ref{figModelStructure} shows a case with only two CM
layers. In the digital world this would correspond, for instance, to
software abstraction layers connected to each other by
simulation/compilation. In the digital world the model CM$^{(2)}$,
which lies directly above the bit-level machine interface model
CM$^{(1)}$, could be the machine instruction level provided by a
microchip manufacturer, or a model written in assembler code. The
highest-level model CM$^{(K)}$ would be expressed in a high-level
programming language or graphical user interface language. In the
digital domain these higher models CM$^{(2)}, \ldots,$ CM$^{(K)}$
would be created by programming experts in the case of programming a
concrete computer, or by theoretical computer scientists in the case
of general formal analyses.

It must be formally specified how input formats $u^{(m)}$ defined in
CM$^{(m)}$ become \emph{encoded} in inputs $u^{(m-1)}$ in the
respective next lower computational model CM$^{(m-1)}$. Conversely,
output formats in the various modeling layers are related to each
other by upward \emph{decoding} rules. The decoding rules can be
formulated in a different formalization language than the formal
definitions of $y^{(m)}$ from $y^{(m-1)}$.

A computational model CM$^{(m)}$ specifies a dynamical system, whose
temporal evolutions --- which we will call 'runs' or 'executions' in
accordance with computer science terminology --- are the
\emph{computations} carried out by CM$^{(m)}$. The broken green arrows
in Figure \ref{figModelStructure} stand for the dynamical laws that
govern the computations. An arrow from $v_i^{(m)}$ to $v_j^{(m)}$
means that the law by which $v_j^{(m)}$ changes its value in time is
co-determined by the values of $v_i^{(m)}$. Each variable $v_i^{(m)}$
has its own local value change law --- we will call it its
\emph{update rule}. Update rules can be deterministic, probabilistic,
or non-deterministic. All these update rules together give the
complete dynamical law for CM$^{(m)}$. When we say that a model
CM$^{(m)}$ is executed, we do not necessarily mean that it is run on a
real physical machine. Demanding that a computational model CM$^{(m)}$
is executable only means that it specifies data transformations which
can be mathematically traced from input to output. All relations
between inputs and outputs which one deems relevant must be
mathematically provable on the basis of CM$^{(m)}$. Most Turing
machines that are described in textbooks are never physically
executed.

\begin{figure}[htb]
  \center 
\includegraphics[width=5cm]{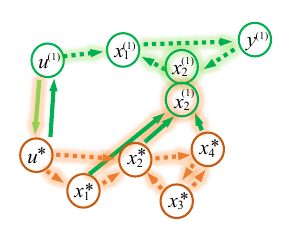}
\caption{Paths in a modeling hierarchy should commute. The graphic
  shows an excerpt from Figure \ref{figModelStructure} with two path
  bands (orange and green shadows) from source $u^{(1)}$ to target
  $x_2^{(1)}$. Such paths must include all loops (here e.g.\
  $x_2^\ast \to x_4^\ast \to x_3^\ast \to x_2^\ast$) and parallels
  (here: $u^\ast \to x_1^\ast \to x_2^\ast$ and $u^\ast \to x_2^\ast$)
  which are touched on the way from the source to the target. The two
  versions of the target determined by the two transformation paths
  (here: the two $x_2^\ast$ nodes with the orange resp.\ green
  shadows) must have approximately the same value. Similar
  commutativity constraints must hold between all levels in the
  hierarchy PM, CM$^{(1)}, \ldots$, CM$^{(K)}$. }
\label{figCommute}
\end{figure}

All broken and solid arrows in and between PM, CM$^{(1)}$, $\ldots,$
CM$^{(K)}$ together should make for a commuting diagram in the
mathematical sense. This entails that if the highest-level output
$y^{(K)}$ is obtained from the input $u^{(K)}$ along two pathways in
our schema --- the first one horizontally using the transformations
within CM$^{(K)}$, the second one first going down vertically through
an encoding sequence until $u^\ast$, then horizontally within PM to
$y^\ast$ and finally decoding upwards again until $y^{(K)}$ is reached
--- the two versions of $y^{(K)}$ thus determined should come out with
(approximately) the same value. The required degree of approximation
is a matter of convention in a given designer/user community. In the
DC world the agreement has to be exact. Also all other horizontal
versus down-horizontal-up path pairs between variables in various
CM$^{(m)}$ should commute (Figure \ref{figCommute}). In the AC theory
literature, commuting diagrams are an important algebraic tool for
formulating consistency conditions between model refinement levels. To
our knowledge, in the physical computing literature this commuting
diagram condition has been first clearly stated by
\citeA{Horsmanetal14}, where a new aspect was that it refers to
equivalence between formal and physical state transformation pathways.

The top layer in Figure \ref{figModelStructure} depicts the real-world
task environment TE where the user of the computing system specifies a
computational task, often in plain natural language, in terms of
initial givens $\alpha^\Omega$ and desired outcomes $\omega^\Omega$,
possibly with subgoals $\beta_i^\Omega$. These items may be connected
by more or less precisely described means-ends conditions (broken
turquoise arrows). The complexity of our real world
will often hinder the identification of well-defined partial states or
observation procedures.

The informal and possibly opaque task description given by an end-user
is captured in a formal task model TM which again \emph{models} the
initial givens, subgoals, desired outcomes and the means-ends
conditions perceived by the user. The modeling relation between the
real task environment and its formal model is much less
well-circumscribed than the modeling relation between PS and PM. There
is no generally followed method that would be equivalent to the
predict-test-refine modeling cycles of the natural science
paradigm. In the digital world, task models TM are mostly specified in
a logic formalism. In classical AI applications they are created by
trained knowledge engineers \cite{Feigenbaum77}.

The formal representation $\alpha, \beta_i, \omega$ of initial givens,
subgoals, and final outcomes is less clear than in the case of the
PS-PM relation. In AI formalisms these representatives may become
extensive logical formulas which only partly describe observable and
identifiable real-world conditions, in contrast to the largely
unequivocal representations of physical measurables in PM. In
statistical formalizations, $\alpha, \beta_i, \omega$ may denote
probability distributions, and in dynamical systems approaches they
may reflect qualitative phenomena like attractors or bifurcations
\cite{Jaeger21a}. Similarly, the specific formal nature of
interrelations and interactions between $\alpha, \beta_i, \omega$
(broken blue arrows) depends on the formal modeling framework that is
used, to be interpreted as causal, correlational, or logical
relations. Task models may or may not be dynamical systems
models. They may or may not be executable, and the typical case is
that they are not. The \emph{operationalization}, which designs the
highest-level computational model CM$^{(K)}$ from TM, in any case
leads to an executable dynamical system model.

There is a wide conceptual and formalism gap between the task model TM
and the highest computational model CM$^{(K)}$. How this gap can be
bridged is one of the most challenging acts in modeling a complete
computing system. In the digital computing domain, this is the
objective of \emph{program verification}, for which a compendium of
subtle and demanding methods has been developed. With regards to a
general theory of modeling computing systems we cannot say much at
present. A core difficulty of connecting TM with CM$^{(K)}$ is that TM
will often only specify the givens and goals without indicating the
effective mechanisms needed to solve the task, while conversely the
formalisms used for computational models CM$^{(m)}$ can codify
mechanical data transformation mechanisms but have no native means to
encode what the data structures mean in terms of task conditions. In
the terminology of computer science this is the distinction between
\emph{declarative} (in TM) versus \emph{procedural} (in CM$^{(m)}$)
specifications.

A naive end-user of a computing system will normally be unaware of
the formalizations TM, PM, CM$^{(1)}$, $\ldots$, CM$^{(K-1)}$ and only
use the highest-level computational model CM$^{(K)}$ in the format of
a high-level programming language or graphical user interface. The
grey arrows in the graphic depict the main interactions of an end-user
with a computing system. He/she knows how to directly \emph{instruct}
the computing system about the task initial conditions $\alpha^\Omega$
in terms of input formats $u^{(K)}$ and how to physically \emph{write}
$u^{(K)}$ values to the physical input states $u^\Psi$, for instance
by hitting a keyboard. Similarly the user must know how to \emph{read}
physical outputs (for instance illuminated pixels on a screen) as
formal computation results $y^{(K)}$, and \emph{interpret} $y^{(K)}$
as real-world task outcomes.

All of these grey-arrow operations cross the ontological border
between abstract formalism and physical reality. Likewise, the two
modeling relations between TE and TM, and between PM and PS, cross
this border. These ontological transitions are not a proper part of a
formal theory computing systems. They represent conditions of its
practical use, human judgement and engineering craftmanship. We
consider only the formal layers PM, CM$^{(m)}$ and TM as a model of a
computing system. In \citeA{Jaeger21a} we called the layers PM and
CM$^{(m)}$ ``how-models'': they capture the procedural mechanics in a
computing systems. In contrast, the task model TM is a ``what-model''
in that it specifies the semantics (goals and real-world setting) of a
computation.

Our schema of models of a computing system may look overly complex
compared to the simplicity of the Turing machine model.  Our schema
has, however, a wider scope than the Turing machine model: we aim at
providing a general conceptual coordinate system for discussing all
the modeling levels that are needed to establish a full-scope
engineering discipline of general computing systems, from physical
substrates to `programming' layers to use-cases. The digital theory
canon offers dedicated, rigorously interconnected
formal subtheories for all layers in our schema. It minimally
comprises the theories of Boolean logic, formal languages,
computability, computational complexity, and first-order logic. The
Turing machine model can be positioned at any of the CM$^{(m)}$ levels,
depending on how closely the physical system model PM resembles a
Turing machine. If the hardware is built with a re-writeable 
tape for a memory mechanism (as in some museum exhibits), the
Turing machine model would sit at the lowest-level CM$^{(1)}$. If a
Turing machine is simulated on a PC via a tutorial graphical
interface, the Turing machine model would come out as the
highest-level model CM$^{(K)}$.

We conclude this section with a comment about our commitment to
state-based transformation formalizations in the computational models
CM$^{(m)}$. CS textbooks include various formalisms that specify the
class $\mathcal{C}$ of 'computable' functions
$f: \mathbb{Z} \to \mathbb{Z}$ by non-deterministic
operations. Examples are the classical, pre-Turing specification of
the recursive functions, lambda calculus, type-0 grammars and
other non-deterministic term rewriting formalisms, including sets of
inference rules for first-order logic. However, when any of these
formalisms is to be used for 'effective' computing, the inherent
non-determinism in these specifications of $\mathcal{C}$ becomes
changed into deterministic processing mechanisms by adding
constraints, for example by requiring a specific evaluation order for
lambda calculus (which leads to functional programming languages) or
heuristic search restriction rules (for term rewriting systems and
logic formalisms, leading to automata realizations of grammars or
implementations of the PROLOG logic programming language). These
derived procedural models are then state-based, and only they can be
'executed' by machines. As far as we can see, in the field of CS only
deterministic machines are used in practice. In principle it is also
an option to add constraints which change such non-determistic models
of $\mathcal{C}$ to probabilistic models, which would lead to
physically executable, state-based, stochastic input-output
transformations. This avenue seems not to be exploited in mainstream
CS. Again, such probabilistic models would be state-based. Since our
main objective concerns physically executable models of 'computing',
we opt for a state-based, procedural view on
computational models CM$^{(m)}$, in a physics/engineering mindset.

\section{Algorithmic and cybernetic theory hierarchies}
\label{secIntroCybernetic} 

\begin{figure}[htb]
  \center
  \includegraphics[width=11cm]{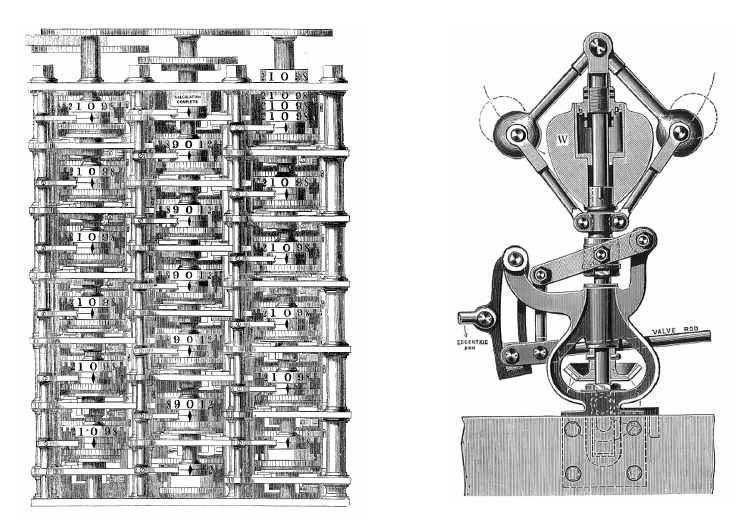}
  \caption{Left: The paradigmatic algorithmic computing machine:
    Charles Babbage's difference engine No. 1 (Harper's new monthly
    magazine, December 1864). Right: The paradigmatic cybernetic
    information processing machine: the centrifugal governor (Drawing
    from W. Ripper: Heat Engines. Longman, London 1909). Both figures
    found in Wikemedia Commons.  }
  \label{figClassicSystems}
\end{figure}

As we will soon explain, in our fluent computing approach toward a GFT
we combine ideas that originated in algorithmic computing (AC) with ideas
from the cybernetic computing (CC) world.  In this section we describe
relevant aspects of both views in more detail, contrasting a schematic
algorithmic model hierarchy with a cybernetic one. Figure
\ref{figClassicSystems} casts a glance at the venerable historical
context of our endeavours.

\begin{figure}[htbp]
  \center
  \includegraphics[width=14cm]{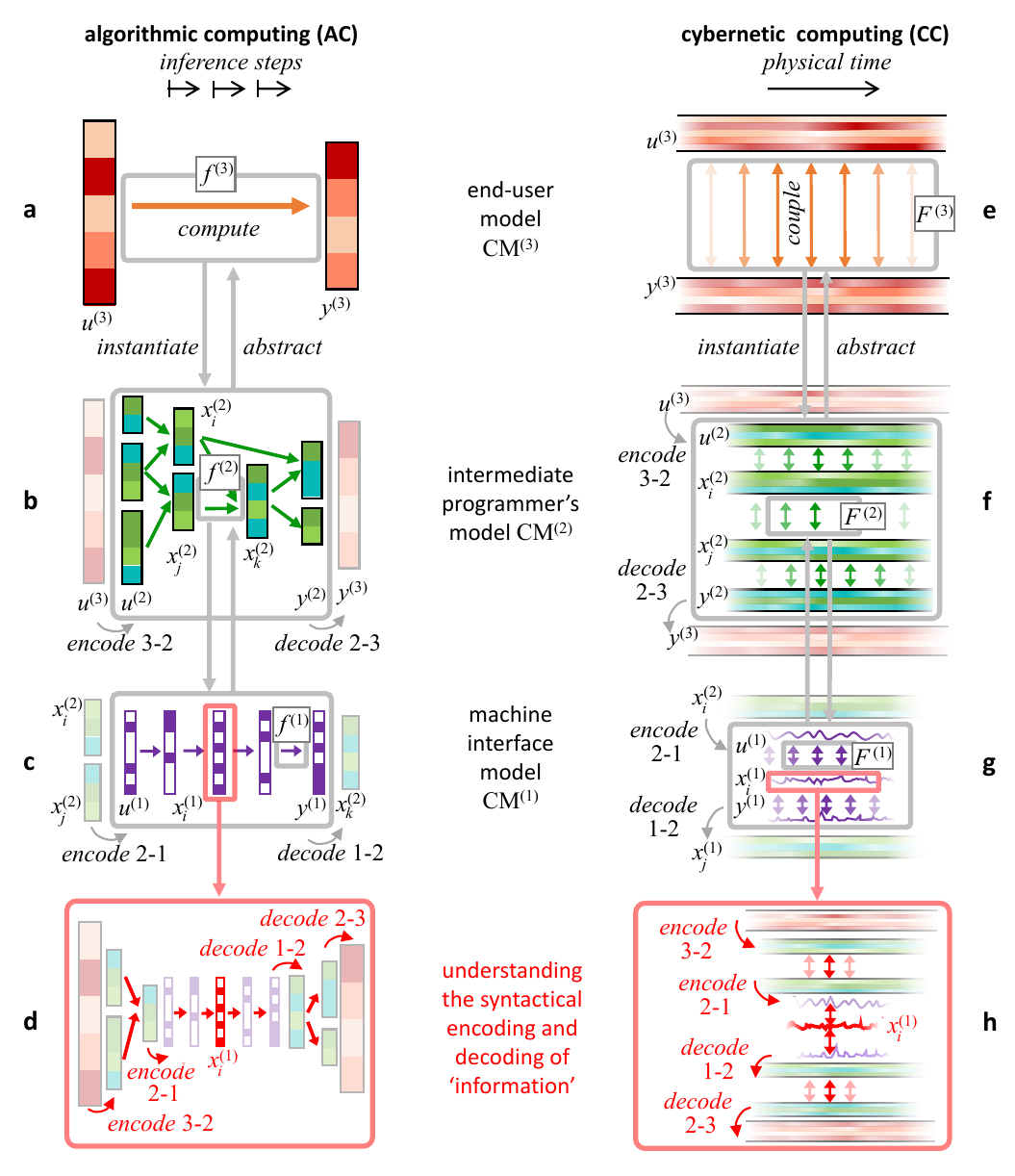}
  \caption{Schematic of algorithmic (left, {\bf a}--{\bf c}) and a
    cybernetic (right, {\bf e}--{\bf g}) modeling a computation on
    three levels. Bottom panels {\bf d}, {\bf h} in red highlight the
    different interpretations of 'information processing'. For
    explanation see text.}
  \label{figAlgCyb}
\end{figure}

The central segment in our generic schema for structuring GFTs (Figure
\ref{figModelStructure}) is the hierarchy of computational models
CM$^{(m)}$.  In Figure \ref{figAlgCyb} we place two such
CM$^{(m)}$-hierarchies side by side, the first one highlighting how
such hierarchies are modeled in an algorithmic spirit (panels
{\bf a}--{\bf c}), while the second one illustrates how we see
cybernetic computing ({\bf e}--{\bf g}; {\bf d} and {\bf h}
will be discussed later). We picture three modeling levels each. At
the top (CM$^{(3)}$) we show a model of a computational process as
it might exist in the mind of an entirely naive end-user, who knows
nothing about the inner workings of the computing machine and only has
a global understanding of the task's input-output transformation. In
the middle (CM$^{(2)}$) we draw a schematic of how this global model
is more concretely instantiated as a computer program by a CS
programmer (left) or as a system configuration by a cybernetic
system engineer (right). At the bottom  we picture a
close-up in CM$^{(1)}$ of a part of the CM$^{(2)}$ model, in which a sub-mechanism
of CM$^{(2)}$ is detailed (compiled, down-engineered) to a level of
detail that is suitable for getting directly mapped to the underlying
hardware.  To make our discussion concrete, we exemplify our
explanations with the elementary algorithmic task of multiplying 6
with 5 on a pocket calculator \cite{Schroeder22}, and the paradigmatic
cybernetic task of regulating the speed of a steam engine with a
centrifugal governor \cite{Maxwell1886}.

We begin with the algorithmic dissection of the 6 $\times$ 5
multiplication task (left part of the figure). Digital computations
are theoretically modeled and practically programmed by breaking down
the overall input-output function $f^{(3)}$ to a low-level machine
interface formalism, through a cascade of increasingly finer-granular
formalisms --- think of compiling a program written in a
high-level programming language like Excel down
to machine-specific assembler code, passing through a series of
programs in languages of intermediate abstraction like Python or C. On
the top level CM$^{(3)}$ in our pocket calculator example, the user
specifies task instances by typing the input string $u^{(3)} =$ {\tt 6
  * 5}, then pushes an 'enter' button and expects the result
$y^{(3)} =$ {\tt 30} to appear in the display.  In our graphic in
panel {\textbf a} the structure of the input {\tt 6 * 5} is symbolized
by the cells in the vertical state bars, and the specific values {\tt
  6}, {\tt *}, {\tt 5} by colors. The user knows that there is a
mathematical function $f^{(3)}$ (the multiplication of integers) which
is evaluated by the calculator, but what happens inside the calculator
of no concern
to a naive end-user. The technical details are taken care
of by the designers of the calculator, who hierarchically break down
the processing of inputs like {\tt 6 * 5} down to a model CM$^{(1)}$,
which matches the available physical elements in the hardware and thus  can be mapped to the logic gates and memory cells of the digital
hardware.

On an intermediate modeling level CM$^{(2)}$ the system designer might
encode the original input $u^{(3)} = $ {\tt 6 * 5} in a binary
representation (for instance {\tt 6} $\mapsto$ {\tt 1 1 0}, {\tt 5} $\mapsto$
{\tt 1 0 1}, {\tt *} $\mapsto$ {\tt binarymult} (leftmost green cell bars
$u^{(2)}$ in panel {\bf b}). This new input encoding is then processed
stepwise with functions $f^{(2)}$ (green arrows), possibly in 
parallel threads, through sequences of intermediate binary
representations $x_i^{(2)}$, until some binary string representation
$y^{(2)}$ of the result is obtained, which then can be decoded into
the top-level representation $y^{(3)} =$ {\tt 30}. One of these
functions $f^{(2)}$ (inner gray rectangle in panel {\textbf b}) could for instance be a binary addition operation
$x_i^{(2)} + x_j^{(2)} = x_k^{(2)}$, corresponding to {\tt binaryadd(1
  1 0, 1 0 0) = 1 0 1 0}.

In a final compilation, this instance of the {\tt binaryadd} operation
might become encoded in a sequence of 8-bit string transformations,
whose outcome $y^{(1)}$ decodes to $x_k^{(2)} =$ {\tt 1 0 1 0}. This
encoding format can be directly mapped to a suitably structured
digital hardware by an experienced engineer. In our schematic diagram
we assume that this level specifies Boolean functions
$f^{(1)}$ (violet arrows in {\bf c}) between 8-bit rewriteable
register models (vertical violet bars). The intermediate level-2 bit
structures $x_i^{(2)}$, $x_j^{(2)}$ become encoded to the 8-bit
register pattern $u^{(1)}$, which serves as the initial state for the
level-1 processing of $f^{(2)}$.

We now turn to our view on hierarchical models of cybernetic
information processing systems ({\bf e} -- {\bf g}). On all
levels $m$, inputs $u^{(m)}$ and outputs $y^{(m)}$ are signal streams
that are continually received / emitted. These signals can
be composed from subsignals --- think of a robot's
overall sensory input which might comprise subsignals from cameras,
touch sensors, the battery status and joint angles. This multi-subsignal
makeup is reflected in Figure \ref{figAlgCyb} by the stripes inside the
$u^{(2)}$, $y^{(2)}$, $u^{(1)}$, $x_i^{(1)}$ and $y^{(1)}$ bands, and
the time-varying subsignal strengths by changing color intensity. The
decomposition into subsignals may be hierarchically continued.  In the
figure only the first-level subsignal structure is shown. 

At the lowest machine interface level $L^{(1)}$ (panel {\bf g}),
input/output signals $u^{(1)}$ and $y^{(1)}$ as well as intermediate
processing signals $x_i^{(1)}$ are modeled as evolving in real time
$t \in \mathbb{R}$.  At higher modeling levels, more abstract
mathematical models $\mathfrak{t}$ of temporal progression may be
used, allowing for increasing uncertainty about precise temporal
localization (explained in more detail in Section 2.4 in
\citeA{JaegerCatthoor23}). We will reserve the word
'signal' for one-dimensional, real-time signals, and use the word
'chronicles' to refer to possibly multi-modal signals that are
formalized with possibly more abstract time models $\mathfrak{t}$.

In our steam engine governor example, the highest-level input
chronicle $u^{(3)}$ would be composed of the measured current engine
speed $s^{(3)}(t)$ and the desired speed $d^{(3)}(t)$. The output
chronicle $y^{(3)}$ is the controlled steam valve setting signal
$p^{(3)}(t)$. The input and output chronicles are continually
connected by a coupling law $F^{(3)}$, in our governor example for
instance by the simple proportional control rule
$\dot{p}^{(3)} = K \, (d^{(3)} - s^{(3)})$. While in this example, the
coupling is unidirectional from $u^{(3)}$ to $y^{(3)}$, in general we
admit bidirectional couplings. In the terminology of signals and
systems, we admit autoregressive filters for coupling laws. We note
that the reafference principle in neuroscience \cite{Jekelyetal21}
stipulates that output feedback is common in biological neural
systems.

While in the governor example, the highest-level input
chronicle $u^{(3)}$ is composed of two signals  which are not
further subdivided, in general a chronicle can be composed of
sub-chronicles which in turn can be compositions of 
sub-subchronicles, etc. Chronicles are hierarchically structured
signal objects (deeper nesting detail not visualized in our figure). 

An intermediate CM$^{(2)}$-model would capture the principal structure
and dynamics of the governor, using chronicles that monitor speeds,
forces, angles etc.\ of system components like masses, levers, axes,
joints etc.

On the lowest level CM$^{(1)}$, the dynamical couplings between
CM$^{(2)}$-chronicles are concretized to match the specific design of
a concrete physical governor. For instance, a coupling $F^{(2)}$
between a centrifugal force and a compensating gravitational load
force (inner gray box in panel {\bf f}), which would presumable be
formalized in CM$^{(2)}$ by a differential equation, would be further
resolved in CM$^{(1)}$ to the metric positioning of joints on lever
arms, the weights and sizing and strengths of mechanical parts, etc.,
leading to fine-grained signals $u^{(1)}$, $x^{(1)}$, $y^{(1)}$ like
the current force or velocity components on a specific joint, or ---
in a high-precision model --- measures of temperature or vibration
which have an impact on the part's functioning.

The algorithmic--cybernetic distinction is not a clear-cut either-or
division. An intermediate view on 'computing' is adopted in models of
analog computing \cite{Shannon41, BlumShubSmale89,
  Moore96,BournezPouly21}. Such modeling strategies follow the
algorithmic paradigm in that a high-level function evaluation task is
hierarchically broken down into lower-level flowchart diagrams of
sequential function evaluations like in our diagrams {\bf a}--{\bf
  c}. At the same time they also appear as cybernetic in that their
data structures are composed of analog real numbers, and the functions
$f^{(m)}$ are evaluated through the continuous-time evolution of
differential operators. In another tradition (branching off from CS),
the input-output theory of algorithmic computing has been variously
extended to sequential processing models, for instance through the
concept of interactive computing \cite{WegnerGoldin03} and (symbolic)
stream computing \cite{Endrullisetal20}.

Both digital computers and brains are commonly said to 'process
information'. One might hesitate to call the
centrifugal governor 'information-processing' too, but we see it as a
member of the large family of dynamical input-output transforming
cybernetic systems that also contains brains, and will thus consider
even such simple regulators as
information-processing. 'Information' is an iridescent concept. When
one is serious about characterizing computing machines as
'information-processing', one should have a clear understanding of this
concept. Algorithmic and cybernetic models of computing systems lead
to different conceptualizations of 'information'.

Information is information about something, encoded in something
else. Full accounts of 'information' thus should cover semantics
(aboutness) as well as what one may call syntax (formats of
information-encoding formal objects). The semantics of what is
computed by some machine relates to the external task, or more
precisely, to the understanding that the user of the machine has of
the task. The syntax becomes manifest in the data structure and
processes that are used in some computational model. The eternal
intellectual challenge to come to terms with \emph{meaning} has
spawned a diversity of formalizations created under different
scientific perspectives. Examples are the semi-formal theories of
semiotics in linguistics \cite{Eco79} and the theory of pragmatic
information where information is captured through its impact on the
belief states of a receiver of information \cite{beimGraben06}, to
name only two approaches that are couched in centuries of
philosophical debate.  We have discussed semantical and syntactical
aspects of 'information' in computing systems in more detail elsewhere
\cite{Jaeger21a}, where we called these two sides the 'what' and the
'how' aspects of information processing. Because it is an important
topic, we here give a summary, referring to the panels {\bf d} and
{\bf h} in Figure \ref{figAlgCyb}.

The algorithmic CS theory offers a full account of semantics and
syntax. The syntactical 'how' aspects are covered in CS theory
textbooks about automata and formal languages, and in practically
oriented textbooks about computer architectures and programming
languages. It is mathematically clear how system models at different
abstraction levels CM$^{(m)}$ can be equivalently en- and decoded into
each other. Students practice this in homework exercises, and compiler
or interpreter programs do it mechanically. If one wants to elucidate
what role is played by the creation and subsequent transformation of a
machine-interface data object $x_i^{(1)}$ (Figure \ref{figAlgCyb}{\bf
  d}), the contribution of this nanosecond event to the successful
computation of the user's task input can be precisely determined (and
debugged if needed) by unraveling the cascade of hierarchical
encodings, inference steps, and decodings that lead from the task
input $u^{(3)}$ down to $x_i^{(1)}$ and back up to the task
output. Furthermore, the semantical 'what' aspects are covered by
formal logic, whose semantic theory (called model theory) can
precisely relate all syntactical structures and operations to
set-theoretic structures, which in turn are formal models of a user's
conception of the task. This is taught to students of computer science
in mathematical textbooks of formal logic and practical textbooks of
program verification.

We do not possess a textbook theory of cybernetic computing that comes
anywhere close to the completeness of the CS theory cosmos. There is
an obvious plausible candidate for a formal account of 'information'
in cybernetic computing, namely Shannon information theory. Tools from
this theory are widely used in signal processing and control,
theoretical neuroscience, statistical physics and complex systems
science --- all of these being concerned with dynamical, open systems
that are akin to our view on cybernetic modeling from Figure
\ref{figAlgCyb}. Shannon information theory can quantify the
information in signals (or chronicles) and calculate the information
gain or loss when stochastic signals are transformed (filter operators
$F^{(m)}$ in Figure \ref{figAlgCyb}) or encoded and decoded from one
format to another. Referring to Figure \ref{figAlgCyb}{\bf h}, the
information contribution of a lowest-level signal $x_i^{(1)}$ to the
highest-level task chronicles could be quantified and tracked through
the en/decoding hierarchy and the coupling processes that concern
$x_i^{(1)}$. Shannon information is framed in probability
theory. Methods from probability theory are widely used in signal
processing, when signals are formalized as stochastic processes. Some
models of information processing in machine learning or computational
neuroscience are expressed in terms of operations on probability
distribtions, for instance dynamic Bayesian
networks \cite{Murphy02a}, statistical physics models like the
Boltzmann machine \cite{Ackleyetal85}, or neural sampling models
\cite{Pecevskietal11}. The computational processes in such models are
often based on sampling, which is a genuinely probabilistic way of
'running' such models. \citeA{Jaeger21a} discusses sampling-based
models of computing in some detail. Shannon's theory can naturally
serve as an interpretation of 'information' in such models.

Furthermore, the rooting of Shannon
information theory in probability theory could yield a
semantic interpretation of a user's task model, which would be
formalized as the probability space that generates the input and
output signals. Like the logic-based semantical models of CS,
probability spaces are richly structured sets, formalized directly in
the langauge of set-theory. Textbooks of information theory
\cite{CoverThomas06} unfortunately do not mention or pursue these
probability-theoretical underpinnings, which could lead a way to a
semantic theory for Shannon information.

Although Shannon information theory appears a natural candidate to
understanding 'information' in cybernetic models, we are skeptical
whether it is the ultimate answer. Casting a cybernetic theory firmly
in a framework of probability might be procrustinating, and for
understanding qualitative phenomena in nonlinear dynamical systems
(attractors, bifurcations and more), which may be important for
computing functionality, we have come to the conclusion that novel
formal tools for capturing their 'how' and 'what' are needed
\cite{Jaeger21}.

Summarizing our contrasting of AC versus CC modeling,
we find that online, real-time, brain-like, cybernetic transformations
from input signal streams to output signal streams is in some aspects
similar to, and in other ways fundamentally different from algorithmic
computing. The similarities justify to classify cybernetic computing
as 'computing' in the first place.  We wrap up the differences in a
punchline: \emph{Algorithmic models of computing
  systems capture the processing of structures, while cybernetic
  models describe the structure of processes.}

\section{Big challenges ahead}\label{secChallenges}

Our schema from Figure \ref{figModelStructure} is just that, a
structure schema for theory systems for physical computing systems. It
is not itself a theory. Yet, this schema already allows us to
identify and discuss problems that at present still limit
our abilities to engineer general non-digital computing systems. In
the domain of digital computing, these problems often are non-issues,
but they become fundamental in physical computing. 

\begin{description}
\item[Which physics?] Any given slab of physical material hosts an
  large number of principally observable local states $v^\Psi$ (Figure
  \ref{figModelStructure}) --- infinitely many for all practical
  purposes. They differ in spatial extension (from individual magnetic
  spins and voltages at electronic contact points to extended
  electrical or magnetic fields or the global temperature of the
  entire slab), time constants (from molecular configuration flips to
  ultra-slow reconfiguration of crystal structures), or complexity
  (from a simple capacitor charge to highly organized resonance
  patterns like skyrmions). New kinds of identifiable and measurable
  nanoscale phenomena continue to become discovered as materials
  science progresses. Their number explodes when one considers the
  possible kinds of dynamical interactions between them and their
  susceptibility to external physical inputs. This raises the question
  which local states $v^\Psi$ and their modeling variables $v^\ast$
  are chosen as a basis for the physical model PM, and how to
  fabricate material systems PS that reliably realize this choice. Or,
  asking the reverse question, if one wishes to realize in hardware
  some computational model CM$^{(m)}$, which materials and phenomena
  can support the state constructs $y^{(m)}$ from that model?  While
  today the standard choice for silicon physical substrates is to
  focus on electronic observables, silicon wafers can also be
  processed into computing systems whose mode of operation is
  mechanical \cite{Coulombeetal17, Dubceketal21} or optical
  \cite{Freibergeretal17}. A consequence of this phenomenal openness
  is a lack of guidance for materials scientists as to which sort of
  physical effects, exactly, they should target and optimize in their
  novel computational substrates.

  In the light of this richness of options it is no surprise that it
  is not clear at present which states in biological brains are the
  relevant ones to explain neural information processing: the states
  of individual synapses, dendritic branches, neural potentials,
  activity patterns in neural microcircuits, or extended neural fields
  --- or which combination of these at which levels of resolution? In
  terms of our generic schema in Figure \ref{figModelStructure}: which
  physical model PM of a brain would be needed to obtain all
  cognitively relevant computations by defining hierarchies of
  computational models CM$^{(m)}$ on top of PM?

\item[What's the time?] Computing systems transform input to
  output. On the physical level PS, this happens in physical time,
  which can be measured with physical clocks. Physical system models
  PM typically (maybe necessarily) use the real-valued timeline
  $t[\mbox{sec}] \in \mathbb{R}$ as formal model of time, either
  directly as in ODE models, or in a sequence of sampling intervals
  $(n \, \Delta t[\mbox{sec}])_{n \in \mathbb{N}}$ as in formal models
  of digital signal processing or time-discretized physical models of
  digital microchips. Above the PM modeling level many other formal
  models of time may be chosen, which are increasingly abstracted from
  physical time $t[\mbox{sec}]$. An extreme case of temporal
  abstraction occurs in the Turing machine. The update steps that lead
  from one Turing machine configuration state to the next one are, in
  a deep sense, timeless. All that is left is the strict sequential
  ordering of configurations. This ordering is of a logical rather
  than a temporal nature: Turing devised of his ``machine'' as a model
  of \emph{logical} reasoning steps. He made it a point that these are
  decoupled from physical time: \emph{``It is always possible for the
    computer to break off from his work, to go away and forget all
    about it, and later to come back and go on with it''}
  \cite{Turing36} --- provided that ``the computer'' left a written
  note to remind him later at which stadium in the computation he took
  a break. The formal model of time in a Turing machine reduces to a
  sequence $s_0 \Rightarrow s_1 \Rightarrow \ldots$ of update steps
  that is ordered purely by logical implication. Thus, in a formal
  model of a computing system, which has the Turing machine in one of
  its levels CM$^{(m)}$, the question arises how Turing state
  variables $v^{(m)}_i$ and their logical progression links (broken
  green arrows in Figure \ref{figAlgCyb}) can be formally defined on
  the basis of a physical system model PM that uses physical time
  $t[\mbox{sec}]$.  If one starts to seriously think about the problem
  of formalizing time and deriving formal hierarchies of increasingly
  abstract models of computational progression, one faces a host of
  conceptual and mathematical questions relating to time constants,
  timescales, synchronization, online vs.\ offline processing, models
  of temporal reactivity and memory span hierarchies that awaits a
  systematic treatment.  \citeA{JaegerCatthoor23} draw an initial
  systematic chart of models of temporality (called ``modes of
  progression'' there) in computational models, and survey about
  twenty partial models and formal transformations between models of
  temporality that have been proposed in digital computer science,
  computational neuroscience, dynamical systems theory and machine
  learning. In the present article we will use the symbol
  $\mathfrak{t}$ when we want to refer to any mathematical model of
  temporal progression.

  We mention in passing that the problem of modeling temporality comes
  with a twin problem of abstraction hierarchies of
  \emph{spatial} scales that spans from metric physical space to
  abstract topological spaces.

\item[How fast does time go by?]  Digital computers employ two basic
  physical memory mechanisms by which bit values are preserved over
  time: the fixed-point attractor stabilization of a high or low
  voltage in the logic gate circuit electronics --- this has a memory
  span of one clock cycle; and the unbounded-duration storage of bit
  values in non-volatile memory devices. By writing to and reading 
  from memory devices, digital computer programs can perfectly
  preserve bit values over any memory span between a clock cycle and
  infinity. This is of obvious importance for executing symbolic
  program code: a value that has been assigned to a variable must be
  reliably retrievable through that variable at any later time where
  it will be needed.

  This is in contrast to biological brains where a plethora of
  physiological mechanisms and anatomical structures are dedicated to
  the encoding, transporting through time, and decoding memory
  items. These mechanisms operate physically in a wide spectrum of
  timescales in a fine-tuned concert. The brain's owner can recall
  previous experiences through a seamless range from milliseconds to
  lifetime without being aware of multitude of neural mechanisms that
  are invoked.  In unconventional computing hardware --- analog
  neuromorphic microchips in particular --- perfectly non-volatile
  memory devices are often not available. Hardware engineering under
  fabricability constraints typically can supply only a small choice
  of physical effects for memory devices and memory subsystems, each
  of which is volatile and supports memory spans in a limited timescale
  interval only. Very long timescales are especially difficult to
  realize.  This leaves wide gaps between the physically available
  timescales, which need to be filled by computational 'tricks' if one
  wishes widely applicable computing solutions.  It is a rich topic
  with roots and repercussions in physics, mathematics, cognitive and
  neuroscience, microchip technologies and philosophy. We cannot give
  an appropriate summary here and must refer to the
  extensive survey and analysis of the timescales challenge for
  physical computing that we gave in \citeA{JaegerCatthoor23}.

\item[Where can I find you? How do you look like?] In digital
  computing, variables $v^{(m)}$ in a model CM$^{(m)}$ can be defined
  through symbolic composites made from variables $v^{(m-1)}$ in the
  modeling level below. These definitions are exactly reversible: the
  composite structures in level $m-1$ that correspond to $v^{(m)}$ can
  be uniquely and effectively constructed from $v^{(m)}$. When
  CM$^{(m)}$ and CM$^{(m-1)}$ are both computer programs written in
  different programming languages, the downward construction is done
  by a compiler and the upward definition is done by abstraction
  operations that are implicit in the compiler design, and could in
  principle be made explicit by a human analyst who tries to decompile
  the lower-level code. When CM$^{(m)}$ and CM$^{(m-1)}$ are abstract
  models as found in computer science textbooks (like a Turing machine
  or a RAM machine model), theorists demand that there is an effective
  bi-simulation relation between the two models. Thus, by
  transitivity, a high-level model CM$^{(m)}$ can be perfectly and
  effectively re-expressed as a machine-interface level model
  CM$^{(1)}$. One step further down, the ontological divide between
  the formal model CM$^{(1)}$ and the physical computing system PS
  must be bridged. Hardware engineers on one side of this divide must
  devise of ways how the lowest-level abstract variables $v^{(1)}$,
  passing through a physical system model PM, are mapped to different
  (ensembles of) physical bit objects in the underlying machine. Let
  us call this the physical addressing problem. Digital hardware is
  structurally organized to enable an effective identification and
  localization of physical bit objects. For instance, magnetic bit
  cells on a harddrive are physically arranged in tracks and sectors
  of known bit length; the electronic bit memory devices in a RAM or
  buffers are arranged and wired up in lines or matrices; the cells on
  a Turing tape are ordered in a discrete sequence.

  In unconventional hardware, physical addressing may become a
  challenge when the structure of the underlying hardware system is
  incompletely known. This may happen in trainable
  hardware that configures its physical structure by an initial
  formatting self-organization process, or due to device mismatch, or
  to unconventional hardware that changes its structure during
  operations. This is certainly the case for biological brains
  \cite{HoltmaatSvoboda09}.  The customary conditions in digital
  computing, which enable the
  mapping of variables $v^{(1)}$ to specific locations or structures
  in the material substrate, then partially or completely get lost.
  Connecting CM$^{(1)}$ with PS will require new ideas.

\item[How many is a computer?] Digital computing machines come in
  different degrees of universality. PC notebooks, smartphones and
  many embedded microchips can be programmed to solve any Turing-solvable computing task provided that they can allocate enough
  physical memory. Theoretical computer science abstracts such
  general-purpose machines as \emph{universal} Turing machines. On the
  other end, application-specific microchips (like stabilization
  controllers in model helicopters and myriads others) can serve only
  one hardwired kind of task, offering limited configuration
  options for a few operational parameters. The formal model of such
  systems are non-universal Turing machines. Outside this digital
  perspective one finds numerous other fashions how a given physical
  computing system can be made to serve a wider or smaller range of
  functionalities.  The neural network models which can be realized on
  a neuromorphic microchip (digital or analog) can be trained for a
  large and important but bounded class of tasks by providing
  different training data. Furthermore one may think of hardware
  systems whose very physical structure grows and adapts during their
  use history, enabling a continual expansion (or forgetting) of their
  task repertoire --- biological brains being the role model, but
  maybe the internet can also be seen in this light.

  Programming, configuring, training, adapting, growing, calibrating,
  stabilizing, annealing, evolving: all of these terms point to the
  question of which tasks a physical computing system can be made to
  serve, by which system designs this can be achieved, and how to
  characterize hierarchies of computational power or generality.

  Our proposed schema for modeling computing systems offers some entry
  points to discuss and categorize this wealth of options.  One
  modeling option we have already mentioned (``which physics?''):
  given a physical system, one has much freedom to choose which of its
  states to use in the physical system model PM. Different choices
  give different bases for ultimately enabled task functionalities.

  When it comes to an efficient development of computing solutions, the
  physical system model PM will however typically be a primary given,
  and even the lowest-level computational model CM$^{(1)}$ will often
  be fixed. In practical engineering of computing systems, PS, PM and
  CM$^{(1)}$ will often be co-developed in close interaction between
  materials physicists, fabrication engineers and microsystem
  architecture specialists.  It is thus of interest to consider
  scenarios where the primary choice to make is to devise of a
  base-level computational model CM$^{(1)}$, assuming that hardware
  engineers can realize it. In the digital world one knows how to
  design and analyse base models CM$^{(1)}$ (direct machine
  interfacing) and CM$^{(2)}$ (machine instruction sets) with regards
  to desired task flexibility, ultimately using the theory of
  (non)universal Turing machines. It would be a great achievement to
  find generic formalisms for base-level computational models
  CM$^{(1)}$ which allow us to formally characterize different degrees
  and sorts of computational universality achievable in higher-level
  computational models CM$^{(m)}$, given specific instances of
  CM$^{(1)}$ formulated in that generic formalism.

\item[From developmental neuroscience to developmental physics?] When
  a human infant grows up, its brain grows and reorganizes by deleting
  and adding anatomical substance and connections and signal
  pathways. In contrast, engineered tools (hammers, fridges, bridges,
  airplanes, digital computers) do not autonomously re-organize their
  physical substance and structure after fabrication in order to adapt
  to new applications or to acquire entirely new
  functionalities. Engineers attempt to design optimal solutions for
  specific tasks, and when a system design is deemed good enough for
  commercial deployment, the designing stops and the
  systems of that design are built and sold. When the designed systems
  fail to meet demands or new task demands arise, the engineers resume
  work, extend and improve their design, and new systems are
  built. Sometimes it is possible to upgrade an engineered system by
  adding components --- this holds especially for computers, for which
  one can add memory chips or plug in external devices. However, we
  would not call such additive extensions 'growth' or 'development' of
  the same sort as it happens in human infants, because such plug-in
  extensions are not happening autonomously, and because the original
  core system does not physically re-organize, and because the new
  functionalities that become enabled by the added components must be
  pre-conceived in the original design.

  When one takes the motto 'learning from the brain' seriously, and
  when one opens the engineering mind to physical substrates that can
  substantially re-organize themselves (for instance in organic
  chemistry or by re-crystallization after heating and cooling
  anorganic materials), a veritable new challenge for theory-building
  appears on the horizon. How can one formally model
  information-processing systems whose physics is developing in
  response to changing tasks? We do not touch this question in the
  present article. Our general schema for physical computing systems
  theories (Section \ref{secStrucPhysTheory}) remains traditional in
  that it presupposes a fixed physical hardware basis. Generalizing
  the picture drawn in Figure \ref{figModelStructure} to scenarios of
  autonomously task-adapting hardware would require us to  embed that
  schematic into a meta theory loop that cyclically connects
  physical models PM with task models TM. We have not begun to think
  this out.

\item[Combining theoretical modeling with practical engineering.]  It
  is not enough to find a way to model physical computing systems in
  an intellectually satisfying manner. A practically relevant theory
  for physical computing must enable concrete system design. Let us
  discuss this challenge of combining theoretical insight with
  practical engineering support in a little more detail. We find
  that there is a tension between the two.

  We begin with a look at neuroscience modeling. Biological evolution
  is apt to find and exploit any physiological-anatomical mechanism
  that adds competitive advantage. Brains appear as \emph{``giant
    'bags of tricks' ''} which integrate \emph{``a huge diversity of
    specialized and baroque mechanisms''} \cite{Adolphs15} into a
  seamless whole. Neuroscientists attempt to understand brains on
  increasingly abstract and integrative modeling levels
  \cite{Gerstneretal12}, from the microscopic biochemistry of synapses
  to global neural architectures needed for learning navigation
  maps. Explaining how the phenomena described on some level of
  abstraction arise from the finer-grained dynamics characterized on
  the level below often amounts to major scientific innovations. For
  instance, the Hodgkin-Huxley model of a neuron
  \cite{HodgkinHuxley52} abstracts from a modeling layer of
  electrochemical processes and calls upon mathematical tools from
  electrical circuit theory; while on the next level of small neural
  circuits, collective voting phenomena may be explained by
  abstracting Hodgkin-Huxley neurons to leaky-integrator point neurons
  and using tools from nonlinear dynamical systems
  \cite{Ermentrout92}. These ad hoc examples illustrate a general
  condition in theoretical neuroscience: the price that is paid for
  trying to understand the phenomenal richness of brains is an
  diversity of modeling methods at different levels of
  abstraction. Each of these modeling methods is insightful, but they
  do not merge into a unified method that could be used by engineers
  to ``learn from the brain'' how to design neuromorphic or other
  unconventional computing systems. It seems that when one tries to
  model complex physical systems in all their richness on many levels
  of abstraction, the price that one has to pay is an 
  incommensurability of the theories at different levels.

  In contrast, multilevel hierarchical modeling of algorithmic
  computing processes is done with one single background theory that
  covers all phenomena within any modeling level as well as the exact
  translations between adjacent levels. This theory is mathematically
  rigorous, fits in a single canoncial textbook
  \cite{HopcroftMotwaniUllman06} whose contents is identically taught
  to CS students worldwide, and lets an end-user of a pocket
  calculator be assured that his/her understanding of arithmetics
  becomes exactly realized by the bit-switching mechanics of his/her
  amazing little machine. The price paid is that digital machines can
  exploit only a single kind of physical phenomenon, namely bistable
  switching --- a constraint that can be seen as the root cause for
  the problematic energy footprint of digital technologies.

  The challenge of combining multi-abstraction modeling of physical
  systems with practical engineering demands is a very hard one. It is
  also the most comprehensively embracing one in our list in this
  section. Many approaches have been tried, as witnessed by our
  long list in the introduction section.  However, we find that none
  of the proposals that we are aware of fully meets the twofold demand
  of openness to a broad spectrum of physical mechanisms and unifying
  engineering transparency across modeling levels.

\end{description}

\section{Fluent computing} \label{secFluentComp}

In the remainder of this article we outline our strategy for
developing a formal theory of physical computing. Our aim is to
reconcile the two seemingly conflicting modeling demands of capturing
general physical systems with their open-ended phenomenology on the
one hand, and of enabling practical system engineering on the
other. Our strategy is to merge modeling principles that originate in
algorithmic and in cybernetic modeling, respectively. From AC we adopt
the hierarchically compositional structuring of data structures and
processes, which is a crucial enabler for systematic engineering. From
CC we take the perspective to view computing systems as continually
operating dynamical systems, which enables us to model information
processing as the evolution of a physical system.  Our key rationale
for working out this strategy is to start from physical dynamical
phenomena and model computing systems in a hierarchy of increasingly
abstracted dynamical systems models, starting from a
physics-interfacing modeling level CM$^{(1)}$. Our name for such formal
model hierarchies is fluent computing (FC). This naming is motivated
by Isaac Newton's wording, who called continuously varying quantities
``fluentes'' in his (Latin) treatise \cite{Newton1669Colson1786} on
calculus; they are now commonly referred to by their English name
``fluents'' in the history of mathematics literature.

\subsection{A phenomenon is what can be
  observed}\label{secObserverIntro}

All theories of physics are about phenomena that can be observed
(measured, detected, sensed) --- at least in principle, and possibly
indirectly. In the followship of physics, we want to set up our FC
theory such that its state variables $v^{(m)}$ can be understood as
denoting observers.

\begin{center}
  \fbox{\parbox{14.5cm}{Our lead intuition about computational model
      variables $v^{(m)}$ is to understand them as representing formal
      \emph{observers} of ultimately physical-level phenomena. While
      we use the generic term 'observer', it may help the reader's
      intuition to think of them as abstractions of sensors,
      detectors, measurement apparatuses, or human observers. This is
      a fundamental choice of metaphor. It enables us to cast theories
      of computing systems as natural science theories, which are
      anchored in the concept of observables. Furthermore, it gives us
      the freedom to consider any physical phenomenon that is
      observable as a candidate for computational uses.}}
\end{center}

An observer responds to the incoming signals by creating a response
signal. For instance, an old-fashioned analog voltmeter responds to a
voltage input by a motion of the indicator needle. Following the
cybernetic view, we cast observing as a temporal process whose
collected observation responses are timeseries objects. The time axis
of these timeseries may be formalized with modes of progression that
abstract from physical time $t \in \mathbb{R}$, as for instance when
one uses integer timesteps $n \in \mathbb{N}$. We use the symbol
$\mathfrak{t}$ for general modes of progression, and call the recorded
timeseries objects chronicles. We will speak of 'observers' when we
mean the objects denoted by model variables $v^{(m)}$. The object that
is being observed will be called the 'source' of the observation.

In more detail, we spell out the observer concept in the following
way. An observer $v^{(m)}$ reacts to a specific kind of stimulus with an
\emph{activation} response (think of the activation of a visual
feature-detecting neuron or the readings of a voltmeter). This
activation $a_{v^{(m)}}$ may continuously change in time. We admit
only positive or zero activation (no negative activations). In this
decision we follow the leads of biology (neurons cannot be negatively
activated; they can only be inhibited toward zero activation) and the
intuition of interpreting activation as signal energy (energy is
non-negative). We have mentioned that we foresee the introduction
of relaxed models $\mathfrak{t}$ of temporal progression, which
reflect a loss of precision compared to physical time $t$. Similarly,
we foresee that relaxed models $\mathfrak{a}$ of real-number 
activations $a$ will be needed, with the latter possibly being
appropriate only in the lowest modeling level CM$^{(1)}$. The general
format of an activation at some time would thus be
$\mathfrak{a}_{v^{(m)}}(\mathfrak{t})$.

The 'specific stimulus' part is harder to grasp.  We call the specific
kind of stimulus to which the observer is responsive, the
\emph{quality} of the observer. However, one cannot exhaustively
characterize what a measurement apparatus responds to. Consider a
thermometer. While a thermometer is engineered to specifically react
to temperature, it will also be sensitive to other physical effects.
For instance, depending on its design, it will also react (if only
slightly) to ambient pressure, vibration or radiation. In
neuroscience, attempts to characterize what exactly a neuron in a
brain's sensory processing pathways responds to remains a conundrum
\cite{SaalBensmaia14}.  We do not want to become entangled in this
difficult question. Whatever a observer reacts to, we will call the
quality of the observer, and we specify this quality by specifying the
observer itself. While the activation value of a observer varies in
time, its defining quality is unchangeable.

In our proposal, the usual concept of a `value' of a computational
variable splits into two components: an activation and a
quality. Activations are one-dimensional scalars whose temporal
trajectories $\mathfrak{a}_{v^{(m)}}(\mathfrak{t})$ document a
computational process. A trace of a computational 'run' of a model
CM$^{(m)}$ is given by a documentation of these activation
histories. Qualities define 'what' is recorded in a variable's
activation trace. They constitute the dynamical laws that govern the
evolution of a 'run', but they are not documented in a trace of a
model's execution.

Observers can be composed of sub-observers, and sub-observers can
again be compositional objects, etc. For example, a retina can be
defined to be composed of its photoreceptor cells, or a safety warning
sensor on a fuel tank might be combined from a pressure and a
temperature sensor. A plausible composition operator for retina
observers would bind the photoreceptor cells through a specification
of their spatial arrangement, while the pressure and temperature
sensor values might be bound together by multiplication. Many
mathematical operations may serve as composition operators. We remain
open with regards to composition operations; different choices may
lead to different FC varieties. Further below we propose a set of
necessary but not sufficient formal properties for composing
observers, as a starter kit for formalizing an FC theory.

Measuring a physical observable needs time. It is a central insight
from physics and signal processing that measurements of physical
observables, like voltages or frequencies, cannot be measured, and are
not even defined, instantaneously at zero-duration time points. The
measured values become more precisely determined when the source is
observed over longer duration, but prolonged measurement integration
also blurs fine-grained, fast temporal
changes. Duration-precision-reactivity tradeoffs will be a constant
companion for theorizing about physical computing, and this is another
reason why we believe that the continual dynamics view of cybernetic
modeling is appropriate for physical computing theories.  In the
platonic abstractions of mathematical thinking, and by consequence in
the Turing machine model and algorithmic modeling, there is no place
for considering duration for determining values of
variables. Therefore, again by inheritance from Turing, AC formalisms
for computational models CM$^{(m)}$ have no means to express timing
constraints for determining values of computational variables.

During the time that is taken to make an observation, the source must
remain identifiable -- and it must remain existant in the first place.
While this is no problem in the digital world, where transistors
obviously live much longer than the clock cycle time that defines the
observation duration, the separation of soure lifetime versus
measurement timescales is not a priori granted in general computing
systems \cite{JaegerCatthoor23}. If, for instance, one wants to
observe properties of fast-decaying electric fields in order to use
them as activation values for some $v^{(m)}$, one is constrained to
only such properties whose measurement time is not longer than the
field's lifetime. We remark that in neuroscience, short-term physical
reconfiguration is a challenge to understanding how neural information
processing mechanisms can be stably identified across extended
durations, which  has become a subject of dedicated research
\cite{Gallegoetal20}.

Observers can have memory, or to say the same thing in other words,
they can be dynamical systems that have state. Their current
activation response may depend on the history of what they have
observed before. In simple cases, this amounts to some degree of
latency needed before the observer's response settles --- think of an
analog voltmeter, whose metal pointer needle takes some little time to
swing to the right volt number on the scale, after the electrodes have
made contact with a new volt source, due to mechanical and
electromagnetic inertia.  In more complex cases, the current
activation response can result from an involved long-term integration
of earlier signal input --- at an extreme end, think of a human who,
while reading a novel (= observing the text signal), integrates the
strange things that are being related in the story with his or her
world knowledge and previous life experiences. However, also
state-free observers that respond immediately to the current input are
possible. 

In the language of signal processing, observers that transform a
previous input signal history into a current response signal, are
called causal filters. The mathematical theory of signal processing
knows of two ways for formalizing causal filters. The (historically)
first way is to characterize such a filter directly by a function $f$
from input histories to the current output signal values, that is one
has $f((\mathbf{u}(t'))_{t' \leq t}) = \mathbf{y}(t)$, where
$\mathbf{u}(t)$ is the input signal and $\mathbf{y}(t)$ is the output
signal at time $t$.  The second way is to characterize the operation
of such a filter by an internal state vector $\mathbf{x}(t)$, which
evolves according to a law
$\dot{\mathbf{x}}(t) = g(\mathbf{x}(t), \mathbf{u}(t))$. The output
$\mathbf{y}(t)$ is obtained by an output function
$\mathbf{y}(t)= h(\mathbf{x}(t))$ from the current state. The two
views yield equivalent filter operations when the current state
$\mathbf{x}(t)$ is uniquely determined by the previous input history,
that is if there exists a function
$\tilde{f}((\mathbf{u}(t'))_{t' \leq t}) = \mathbf{x}(t)$. This
condition is both fundamental and nontrivial. It has been intensely
studied, under the headlines 'echo state property' or 'fading memory',
in the reservoir computing field
\cite{Jaeger01a,MaassETAL01a,Yildizetal12,ManjunathJaeger13,
  GrigoryevaOrtega18a,GononGrigoryevaOrtega20a}.

Observers $u^{(m)}$ which represent input streams are special in the
  sense that they never have memory. 

  Sometimes one will wish to have observers $v^{(m)}$ that can trace a
  phenomenon while it changes its physical or formal character. For
  instance, one may wish an observer $v^{(1)}$ in the machine
  interfacing model CM$^{(1)}$ to trace a pulse-like signal that
  travels from some location A to another location B, like a neural
  spike traveling from one neuron to another; or like a digital bit
  pulse that is created through electromagnetic effects in a magnetic
  memory device but then travels purely electrically along a wire; or
  a spike which changes its propagation physics from electrical to
  electrochemical to electrical when it hops across a gap in the
  axonal myelin sheath, and which changes from electrical to
  biochemical in the destination synapse. In our view of identifying
  observers with their quality, this leads to a modeling problem: the
  observer's quality is the specific characteristic of 'what the
  observer can see'. When the phenomenon that one wishes to trace
  changes its physical characteristic, the observer should be able to
  keep track across physical sources of different kinds, which at face
  falue apparently requires a sequencing of observers of different
  qualities. Formal solutions to this problem could be temporal
  chaining of observers with different qualities; or defining
  observers whose quality lets them respond to sources of different
  physical kinds; or to make qualities time-varying concepts; or
  exploiting an observer's internal state to let it adapt to changing
  characteristics in the input. In our FC proposal we opt for a
  dynamical re-combination of observers into variable compounds (see
  below).

\subsection{Observers bind and couple with observers within a
  model}\label{subsecObsCouple}

In this subsection we explain the organization of the structures and
processes within an FC model CM$^{(m)}$. We describe how a
hierarchical interaction architecture of observer objects $v ^{(m)}$
is explicitly governed by a mechanism that we call 'binding', and how
from this organization of interaction by binding hierarchies other
interaction patterns emerge, which we call 'coupling'. Coupling
phenomena  are only implicit in  CM$^{(m)}$
and can only be discovered by an analysis of the model dynamics.

Coupling and binding refer to phenomena and mechanisms \emph{within} a
model CM$^{(m)}$. We will discuss relations \emph{between} subsequent
models in the modeling hierarchy CM$^{(1)}, \ldots,$ CM$^{(K)}$ in the
next subsection.

The words 'coupling' and 'binding' are used in many ways, sometimes
interchangeably. To prepare the grounds for our exposition, we take a
brief look at some of the uses of these words in the wider complex
systems sciences.

Physical computing machines are certainly complex systems. Now
'complex system' is not a particularly well-defined concept, and the
field that calls itself 'complex systems science' is itself a complex
merge of concepts and methods from almost all natural and social
social sciences \cite{Thurneretal18_chap1}. However, across the many
perspectives on 'complex systems' in this
field, a common denominator is to model complex systems as being
hierarchically organized in nested structures and processes. Modeling
how smaller subsystems 'couple' or 'bind' into larger compound
subsystems was a core objective for the field from its beginnings
\cite{Simon62}. For our FC proposal, we follow this lead and organize
the architecture of models CM$^{(m)}$ around principles of
hierarchical organization. 

Many system models in the natural sciences and engineering are
expressed through systems of coupled ordinary differential equations
(ODEs) of the form $\dot{x}_i = F_i(x_1, \ldots, x_n)$, where the
specific form of the function $F_i$ determines how strongly
interaction partners $x_j$ influence the evolution of
$x_i$. 'Coupling' here is a technical term, which simply means that
some state variable $x_j$ appears as argument in the right-hand side
of the ODE for some other state variable $x_i$.  This format naturally
captures uni- and bidirectional couplings and self-coupling. Another
common formal framework for defining couplings between state variables
$x_i, x_j$ is to join them by symmetric, weighted links
$w_{ij} \in \mathbb{R}$ in a network graph, where $w_{ij}$ may be
referred to as 'coupling strength'. This is constitutive for systems
models devised in a spirit of statistical physics, as in Ising models,
Hopfield networks and Boltzmann machines
\cite{MarulloAgliari20}. Generalized network-based models are a main
modeling approach in complex systems science
\cite{Thurneretal18_chap1}. An arsenal of methods has been developed
to identify subsets of state variables which are more strongly coupled
with each other than with other variables, allowing some kind of
hierarchical decomposition of the overall system into interacting
subsystems. The specific goals of such analyses vary, and the explored
methods are not trivial. Important clues about a hierarchical system
organization can be obtained from analysing the static connectivity
matrix in network models, which is already challenging
\cite{Thurneretal18_chap4}, but the task becomes even more difficult
when the hierarchical organization is to be identified in the
dynamical interaction patterns that arise in complex systems. Examples
are the explicit design of hierarchical control architectures
\cite{Albus93}, stability analyses based on contractivity conditions
for subsystems in modular control systems \cite{SlotineLohmiller01},
methods for model order reduction where sub-collections of system
variables are replaced by single new variables by linear algebra
methods \cite{AntoulasSorensen01}, the multiscale spreading dynamics
of perturbations in dynamical systems defined on hierarchical graphs
\cite{Hensetal19}, and many other methods developed in the
mathematical theory of multi-timescale systems \cite{Kuehn15}, or
self-organizing dynamical systems \cite{Haken08}.

A hierarchical organization of formal models appears almost inevitable
when it comes to modeling information-processing systems, too.  All
AC models are based on mechanisms to hierarchically
compose more elementary symbolic data structures into more complex
compound data structures. The simplest case is to bind atomic symbols
into finite sequences ('words'). For instance, the word {\tt 00000011}
(which might represent the number 3 in 8-bit binary representation) is
compounded from {\tt 0} and {\tt 1} symbols. On the other end of the
complexity scale one finds deeply organized data structures that
represent spreadsheet documents or even entire databases. Besides data
structures, data transformation operations are likewise hierarchically
structured. The classical example is the mathematical definition of
recursive functions, which casts every 'computable' function as
hierarchically combined from a few elementary functions, like the
integer successor function
$\sigma: \mathbb{N} \to \mathbb{N}, n \mapsto n+1$ (the recursive
functions turn out to be exactly the functions that can be computed by
Turing machines). Mathematicians and logicians ubiquitously create new
formal objects by composition from previously defined ones. Also, the
wider cognitive science and AI literature describes many phenomena of
conceptual 'chunking', in particular in connection with analyses of
planning, working memory or the organization of percepts
\cite{Lairdetal86,Drescher91,Baddeley03,SchackRitter09}.  In
theoretical neuroscience, the activation dynamics of subsystems and
their activation-induced bindings and unbindings of into composite
associations is central for understanding cognitive information
processing. The 'binding problem' remains a core challenge for
understanding how cognition arises from neural interactions
\cite{Treisman98,Diesmannetal99,Shastri99a,SlotineLohmiller01,Legensteinetal16}.

There is a noteworthy difference between the hierarchical structure in
high-dimensional ODE models in the sciences and engineering on the one
hand, and the hierarchical structure in algorithmic computing models
on the other hand. In the former, the hierarchical structuring is
implicit in the model and has to be brought to the surface by
mathematical analyses that are not part of the model itself. In the
latter, hierarchical data structures and process compositions are
explicit in the model; structure-forming operations are key in
formulating these models in the first place. Furthermore, couplings in
ODE systems are graded (variables and subsystems can be bound together
with different strengths), while in symbolic computer programs and
other algorithmic system models, binding formal components together is
a categorical yes/no decision.  For our strategy to build FC
models, we want to unify these two perspectives on hierarchicity,
because FC models should be interpretable both as modeling physical
systems (like ODE models in the natural sciences) and as modeling
information processing (like AC models do).

We will use two words, 'binding' (for explicit, categorical yes/no
compositions) and 'coupling' (for implicit, graded interaction) to
keep these two perspectives apart. In our proposed FM modeling
approach, they are separate mechanisms. Before we go into detail, we
point out a formal similarity between hierarchical relations within,
and between, computational models CM$^{(m)}$.

We recall that our prime motivation to cast the objects denoted by
model variables $v$ as observers, was to be able to use many sorts of
physical phenomena as basis for 'computing'. This let us set up
computational models CM$^{(m)}$ in analogy to system models in the
natural sciences, where model variables stand for physical
observables. In our proposed hierarchy of models, a higher-level model
CM$^{(m)}$ models physical reality indirectly, by letting its
observers $v^{(m)}$ observe the observers $v^{(m-1)}$ in CM$^{(m-1)}$,
etc., until at the level closest to physics, observers $v^{(1)}$
observe the physical model PM whose state variables, finally, directly
stand for physical observables.

We will allow observers $v^{(m)}$ to observe other observers not only
across adjacent models from CM$^{(m)}$ to CM$^{(m-1)}$, but also
within CM$^{(m)}$. The specification of how some $v^{(m)}$ observes
some $v'^{(m-1)}$ in the model below will be formally identical to the
specification of within-model observations of $v'^{(m)}$ by
$v^{(m)}$. Observer-observee relations will form the basis for
establishing subsystem hierarchies within CM$^{(m)}$, as well as the basis
for ordering models CM$^{(1)}, \ldots,$ CM$^{(K)}$ in a modeling
hierarchy.

We can find a similar formal equivalence between within- and
across-model hierarchies in AC program compilation hierarchies, where
a program CM$^{(m)}$ written in a higher-level programming language is
compiled into a program CM$^{(m-1)}$ in a lower-level programming
language. Often a program is organized in a way that in a preamble a
number of complex data structures and functions are defined and given
names. The body of the program, which 'does the job', then uses only
these named compound modules. This is tantamount to using a
programming language that only employs the defined, named compound
modules, and regard the original defintions in the preamble as compile
instructions to a lower-level language.  Indeed, some higher-level
languages (like Python or Matlab) admit the direct inclusion of code
fragments from a lower-level language like C. Toolbox languages (like
TensorFlow for implementations of machine learning applications)
provide high-level data structures and languages directly, which are
tailored to some application domain and relieve the modeler from the
task to define the needed compound modules himself or herself.

Within-model observation relations will form the basis for a
hierarchical structuring of models CM$^{(m)}$, and we will reserve the
word 'binding' for these model-internal structuring
mechanisms. Across-model observation relations are the basis for model
hierarchies, and we use the word 'abstraction' when we refer to how
CM$^{(m)}$ relates to CM$^{(m-1)}$.  In the remainder of this
subsection we refer to a fixed modeling level $m$, and we will mostly
omit the superscript $\cdot^{(m)}$ for easier reading. We now finally
can give a semi-formal, step-by-step explanation of how we cast
'binding'.

\begin{description}

\item[Scoping observers and observees.] At any time $\mathfrak{t}$ in
  the execution of CM$^{(m)}$, an observer $v$ observes a set
  $\mathcal{B}^\downarrow_v(\mathfrak{t})$ of other observers $v'$ in
  CM$^{(m)}$. An observer $v$ cannot observe itself:
  $v \notin \mathcal{B}^\downarrow_v(\mathfrak{t})$. An observer need
  not observe anything inside CM$^{(m)}$:
  $\mathcal{B}^\downarrow_v(\mathfrak{t})$ can be empty. Conversely,
  $\mathcal{B}^\uparrow_{v'}(\mathfrak{t})$ denotes the set of
  observers that observe $v'$ at time $\mathfrak{t}$. Again, $v'
  \notin \mathcal{B}^\uparrow_{v'}(\mathfrak{t})$ and
  $\mathcal{B}^\uparrow_{v'}(\mathfrak{t}) = \emptyset$ is
  possible. We call $v$ \emph{atomic} (at time $\mathfrak{t}$) when
  $\mathcal{B}^\downarrow_v(\mathfrak{t}) = \emptyset$, and
  \emph{autonomous} when $\mathcal{B}^\uparrow_v(\mathfrak{t}) =
  \emptyset$. 
  
\item[Observation yields an order relation.] We furthermore require
  that observation relations yield an order relation on the set
  $\mathcal{V}(\mathfrak{t})$ of observers $v$ that are instantiated
  in CM$^{(m)}$ at time $\mathfrak{t}$. To make this formal, let
  $v' \lhd_{\mathfrak{t}} v$ denote the condition that $v'$ is
  observed by $v$ at time $\mathfrak{t}$, and let
  $\lhd_{\mathfrak{t}}^\ast$ be the transitive closure of
  $\lhd_{\mathfrak{t}}$. We demand that $\lhd_{\mathfrak{t}}^\ast$ is
  a strict order relation, that is
  $v_1 \lhd_{\mathfrak{t}} v_2 \lhd_{\mathfrak{t}} \ldots
  \lhd_{\mathfrak{t}} v_n$ implies $v_1 \neq v_n$.

\item[Being observed together = being bound.] We say that the elements
  $v' \in \mathcal{B}^\downarrow_v(\mathfrak{t})$ are \emph{bound} by
  $v$ at time $\mathfrak{t}$.  The order $\lhd_{\mathfrak{t}}^\ast$
  thus establishes a \emph{binding hierarchy} at time
  $\mathfrak{t}$. Note that some $v'$ can be bound by different
  $v_i, v_j$ at the same time, that is
  $v' \in \mathcal{B}^\downarrow_{v_i}(\mathfrak{t}) \cap
  \mathcal{B}^\downarrow_{v_j}(\mathfrak{t})$. Some more terminology:
  when $v' \lhd_{\mathfrak{t}} v$, we say that $v'$ is a
  \emph{component} of $v$, and $v$ is a \emph{compound} that is
  \emph{composed} of its components (all at time
  $\mathfrak{t}$). Furthermore, by
  ${\mathcal{B}^\downarrow}^\ast_{v}(\mathfrak{t})$ and
  ${\mathcal{B}^\uparrow}^\ast_{v}(\mathfrak{t})$ we denote the
  transitive hull of ${\mathcal{B}^\downarrow}_{v}(\mathfrak{t})$ and
  ${\mathcal{B}^\uparrow}_{v}(\mathfrak{t})$, that is the set of all
  observers that lie below (resp.\ above) $v$ in the binding
  hierarchy.

\item[Binding is the basis for model organization.] The order
  $\lhd^\ast_{\mathfrak{t}}$ plays a fundamental role in our FC
  proposal. As we will soon see, it yields the overall organization of
  the information processing dynamics within CM$^{(m)}$. With regards
  to its global organization role, it can roughly be compared to the
  definition hierarchies of data types and functions in programs
  written in an imperative programming language; or the definition of
  the objects that interact in the execution of digital computer
  programs written in an object-oriented programming language; or the
  nesting hierarchy of lambda-expressions in a program (itself a
  single lambda-expression) written in a pure functional programming
  language.
  
\item[Time dependence is optional.] The set of observers
  $\mathcal{V}(\mathfrak{t})$ that are present in CM$^{(m)}$ at time
  $\mathfrak{t}$, and their hierarchical binding relations are, in
  general, time-dependent.  We will describe presently how
  $\lhd_{\mathfrak{t}}^\ast$ organizes the interaction dynamics of
  observers within $\mathcal{V}(\mathfrak{t})$. Often, however, a
  model will be set up by the modeler with a fixed set of observers
  and a fixed binding hierarchy, which does not vary during execution
  time. We may say that such models have a \emph{static
    architecture}. Since ultimately (going down from CM$^{(m)}$
  through CM$^{(m)-1}$, $\ldots$, CM$^{1}$, and PM to the physical
  system PS), an FC model CM$^{(m)}$ is cast as observing a physical
  system, models with a static architecture are suitable when the
  physical system itself is not changing over time. Static models are
  certainly easier to define and analyse and map to physical hardware
  than self-modifying models CM$^{(m)}$ whose binding structures
  change during model execution.  Models with such a \emph{dynamical
    architecture} may however be needed when the physical substrate
  underneath changes, for instance through aging, growth, shrinking,
  or other sorts of physical re-organization; or in scenarios where
  different physical phenomena are recruited for computing at
  different times; or when physical phenomena that are ultimately
  observed by some $v$ in CM$^{(m)}$ 'travel' in the physical
  substrate (like electric pulses, solitons, waves or neural spikes)
  and interact with different other phenomena at different times.

  For better readability, we will assume a static architecture in the
  following and drop the reference to times $\mathfrak{t}$ in the
  structure-defining constructs
  $\mathcal{V}, \mathcal{B}^\downarrow_{v}, \mathcal{B}^\uparrow_{v}$
  and $\lhd$,  with the understanding that all of the following
  specifications can also be made time-dependent.

\item[Hidden memory states and visible activations.]  We mentioned in
  Section \ref{secObserverIntro} that observers can have memory, and
  that one possible mathematical method to capture memory is to endow
  observers with an internal state. Here we choose this method, but an
  FC modeler might also use another one. Let
  $\mathbf{s}_v(\mathfrak{t})$ denote the state of $v$ at time
  $\mathfrak{t}$. The state $\mathbf{s}_v(\mathfrak{t})$ is private to
  $v$ and not visible to other observers in CM$^{(m)}$. The only
  signal that an observed $v'$ can reveal to other observers is its
  activation $\mathfrak{a}_{v'}(\mathfrak{t})$.

\item[Memory state updates.] An observer $v$ observes its components
  $v' \in \mathcal{B}^\downarrow_v$ by integrating information from
  their activation signals over time. This is formally expressed
  through a \emph{state update operator} $\sigma_v$, whose action can be
  semi-formally rendered as
  \begin{equation}\label{eStateUpdate}
    \mathbf{s}_v(\mbox{\sf next-}\mathfrak{t}) =
    \sigma_v(\mathbf{s}_v(\mathfrak{t}), (\mathfrak{a}_{v'}(\mathfrak{t}))_{v' \in  \mathcal{B}^\downarrow_v}). 
  \end{equation}

  How the temporal progression
  ``$\mathfrak{t} \mapsto \mbox{\sf next-}\mathfrak{t}$'' is formally
  defined, and what mathematical format the temporal progression
  $\mathfrak{t}$, the memory states, and the update operator
  $\sigma_v$ take, is up to the modeler. For our FC theory-building we
  admit non-deterministic or probabilistic laws (as in many sorts of
  automata models in CS or stochastic process models), and
  time-varying sets of arguments (as in logical inference
  engines). Working out concrete FC models invites the invention of
  new formats for states, activations and temporal progression modes.

\item[Activation updates.] The activation
  $\mathfrak{a}_v(\mathfrak{t})$ is a memoryless function $\alpha_v$
  of the current state of $v$:
  \begin{equation}\label{eActivationUpdate}
    \mathfrak{a}_v(\mathfrak{t}) =
    \alpha_v(\mathbf{s}_v(\mathfrak{t})).  
  \end{equation}

\item[Modulation of dynamics downward through the binding hierarchy.]
  When $v' \lhd v$, we also want to capture 'top-down'
  modulating effects from $v$ to the components $v'$ that are bound by
  it, in analogy to, for instance,
  \begin{itemize}
  \item top-down pathways in neural processing hierarchies, which
    abound in biological brains and serve a multitude of functions
    like attention, setting predictive priors, or modulation of motion
    commands,
  \item passing down arguments in function calling hierarchies in
    AC programs,
  \item passing down subgoals in hierarchical action planning systems
    in robotics and AI,
  \item influencing the collective interaction of multi-particle
    systems by external fields in physics.
  \end{itemize}
  Formally, we frame the activation of the compound observer $v$ as
  a control parameter in the activation update laws of its components
  $v'$. We obtain the following completed version of
  \eqref{eActivationUpdate}:
   \begin{equation}\label{eActivationUpdateComplete}
   \mathfrak{a}_v(\mathfrak{t}) =
    \alpha_v(\mathbf{s}_v(\mathfrak{t}),
    (\mathfrak{a}_{v''}(\mathfrak{t}))_{v'' \in  \mathcal{B}^\uparrow_v}). 
  \end{equation}
  The activations of the compound parents $v''$ of $v$ thus act as
  formal control parameters for $\alpha_v$. Figure
  \ref{figCoupleBind}{\bf a} summarizes the argument structures for
  the state and activation update laws.

  It is a common assumption in complex systems modeling that
  higher-level variables in hierarchical systems are 'slower',
  'spatially more extended', 'coarser', etc., than lower-level
  variables. Hierarchical system models are typically multi-scale
  models. In our FC proposal we do not make assumptions about scale
  separations between the activation dynamics of $v'$ versus $v$ when
  $v' \lhd v$. Binding hierarchies may or may not support a natural
  interpretation of being multi-scale of some sort.

\begin{figure}[htb]
  \center
  \includegraphics[width=11cm]{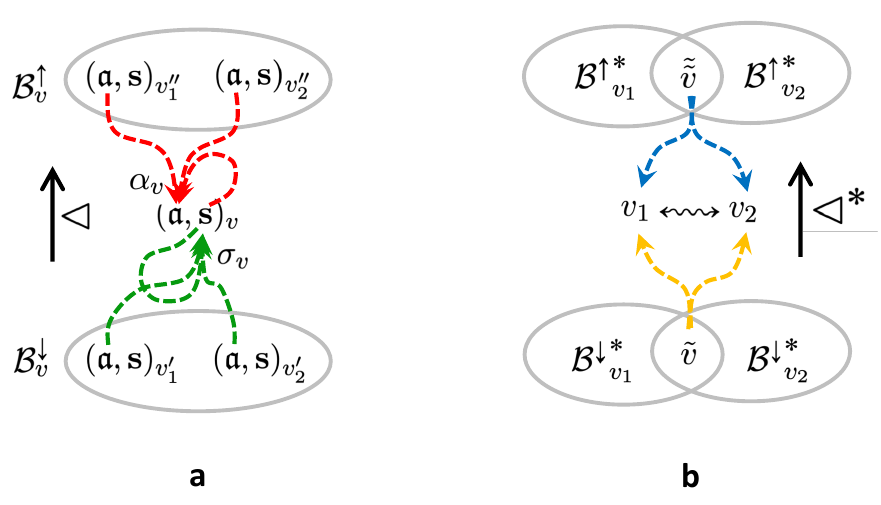}
  \caption{Dependencies for state and activation updates of an
    observer $v$ (in {\bf a}) and for coupling between $v_1$ and $v_2$
    (in {\bf b}). {\bf a:} The activation and state of an observer $v$
    is indicated by $(\mathfrak{a},\mathbf{s})_v$. The state
    $\mathbf{s}$ of $v$ is updated through $\sigma_v$, which lets $v$
    observe the activations of components $v_i'$ of $v$ with memory
    encoded in the previous state $\mathbf{s}$ (green arrows). The
    activation $\mathfrak{a}$ of $v$ is a direct function of the
    memory state $\mathbf{s}$ of $v$, top-down controlled by the
    activations of observers $v''_j$ in which $v$ is a component (red
    arrows). {\bf b:} Two observers $v_1, v_2$ are coupled by shared
    master when there exists some $\tilde{\tilde{v}}$ of which both
    are (transitive) components, which means that both are
    (transitively) controlled by $\tilde{\tilde{v}}$ (blue
    arrows). They are coupled by shared member if there exists some
    $\tilde{v}$ which is a (transitive) component of both, such that
    both $v_1$ and $v_2$ (transitively) observe $\tilde{v}$ (orange
    arrows).  }
  \label{figCoupleBind}
\end{figure}

\item[Bidirectional, asymmetric vertical interaction through binding.]
  We thus have a bidirectional influence between observers $v'$ versus
  $v$ when $v' \lhd v$. The bottom-up influence given by
  \eqref{eStateUpdate} is cast as an observation relation with memory,
  while the top-down influence \eqref{eActivationUpdateComplete} is
  one of memoryless control, as one might say with a very general
  interpretation of the word 'control'.

\item[Lateral, symmetric interaction through coupling.] We now turn to
  the second interaction concept that we want to install in FC models,
  'coupling'. Coupling interactions are not explicitly modeled in
  CM$^{(m)}$ but result indirectly from the dynamics of
  \eqref{eStateUpdate} and \eqref{eActivationUpdateComplete}. We say
  that two observers $v_i, v_j$ are \emph{coupled by shared masters}
  when they share at least one observer $v_k$ that sits higher in the
  binding hierarchy, that is when $v_i, v_j \lhd^\ast v_k$. All
  components $v'$ of a compound observer $v$, and their transitive
  sub-components, are coupled by a shared master, which is $v$. And we
  say that $v_i, v_j$ are \emph{coupled by shared members} when they
  share at least one observer $v_k$ that sits lower in the binding
  hierarchy, that is when $v_k \lhd^\ast v_i, v_j$. In both cases, the
  activation dynamics of $v_i$ and $v_j$ exert some influence on each
  other through the bidirectional binding interactions along the
  binding pathways from $v_i, v_j$ to $v_k$. We write
  $v_i \leftrightsquigarrow v_j$ when $v_i$ and $v_j$ are coupled by
  shared masters, or by shared members, or by both. Figure
  \ref{figCoupleBind}{\bf b} illustrates this.

  Mutual dependencies between observers induced by coupling
  may spread throughout $\mathcal{V}$, indirectly linking observers
  that have neither shared members nor masters: when $v_1
  \leftrightsquigarrow v_2 \leftrightsquigarrow \cdots
  \leftrightsquigarrow v_n$, perturbations to the state or activation
  of $v_1$ will affect the state/activation dynamics of $v_n$. We
  write $v_i \leftrightsquigarrow^\ast v_j$ to denote this indirect
  coupling by transitivity.

  In our diagrams of cybernetic computing (Figure \ref{figAlgCyb}{\bf
    e}-{\bf g}), observers that are bound in composite observers are
  indicated by colored subbands, and the coupling interactions by
  vertical arrows. These diagrams show dynamics in some static
  architecture --- the binding structure does change while time goes
  on. But, note again that the
  existence of couplings may be made temporally varying in dynamical
  architectures.

\item[Analogues of coupling in other models of complex systems.]
  Analogues of coupling by shared members can be found in many complex
  systems, when two subsystems co-own or exchange messages or
  materials generated or located within them. At the most fundamental
  level, this is witnessed by particle interaction models in
  theoretical physics, where forces between particles can be modeled
  as the exchange of force-mediating messager particles
  \cite{GreggJaeger21}. In some AC parallel programming paradigms and
  AI 'blackboard models' of cognitive processing, functions or program
  modules can interact by having simultaneous read/write access to
  information in a shared memory region \cite{Nii86}.

  Analogues of coupling by shared masters occur ubiquitously in
  hierarchical control systems (in biological neural motor control or
  in engineered control systems) when several subsystems receive
  top-down control input from the same superordinate subsystem. This
  enables coordinated action of end-effectors, for instance muscles
  which are co-activated by a central pattern generator in a
  coordinated motor behavior. In physics, ensembles of particles or
  other microscale subsystems may statistically build up a 'field' of
  some sort in a bottom-up way, which in return organizes large-scale
  interaction patterns of the microsystems in a top-down
  direction. Such systems can be modeled with mean-field methods,
  where the 'field' is mathematically characterized as a dynamical
  system constituted by a small number (often even a single) of 'order
  parameters'. Under suitable conditions, the high-dimensional
  interaction of microsystems simplifies and can be characterized by
  the low-dimensional 'field' dynamics in good 
  approximation. In statistical physics, methods of this kind are
  central for explaining phase transitions. In synergetics
  \cite{Haken83}, these methods have been have become generalized and
  adapted to explain 'self-organization' in a wide range of complex
  systems in biology, cognitive science, sociology and economics.

\item[Analysis and design of models is tied to understanding
  coupling.] In our set-up of models CM$^{(m)}$, the 'lateral'
  coupling between model variables $v$ is not explicitly reflected in
  the model. Instead, coupling interactions emerge indirectly through
  information sharing in the binding hierarchy. This is different from
  the 'lateral' coupling between state variables $x_i$ in ODE systems,
  which is directly visible in the system equations. One might say
  that ODE systems are designed around such couplings, and the
  emergence of hierarchical subsystem nestings must be found out
  post-hoc by nontrivial analyses. In FC models this is reversed: The
  hierarchical subsystem structuring is explicitly built into the
  model by its designer, who starts from a conception of this very
  structuring. \emph{Which} variables are indirectly coupled via
  $\leftrightsquigarrow$ is, on the one hand, easily determined from
  the written-down model CM$^{(m)}$ by tracing the transitive member
  and master relations. However, \emph{how} these
  interactions could be best intuitively understood and matched
  against the underlying physical system PS through its physical model
  PM, is equally nontrivial as the induction of hierarchical structure
  in ODE models.  Interactions by transitive coupling certainly can
  lead to decisive, global 'self-organization' (or disruptive 
  'self-disorganization') effects when CM$^{(m)}$ is executed. A
  modeler might want to foresee such important effects and directly
  capture them in CM$^{(m)}$ by introducing master observers that
  observe and control them, making such effects explicit in the model
  from the outset.

  The primacy of hierarchical structuring over emergent
  self-organization phenomena is shared by our FC approach and 
  algorithmic modeling. Classical computer programs and machine
  models likewise start from a hierarchical organization of data
  structures and functions. The ultimate global behavior of a computer
  program, when it is executed, is always open to surprises. Some of
  them cannot be foreseen by any universal prediction method --- the
  most famous example being the undecidability whether a program, once
  started, actually will halt or run endlessly without ever producing
  a result. This brings us back to Turing, who proved this!

\item[Execution of a model.] A model CM$^{(m)}$ must be
  'executable'. This can mean several things to mathematicians, computer
  scientists, or physicists:
  \begin{itemize}
  \item mathematically deriving how an input signal is transformed to
    an output signal as time progresses;
  \item simulating 'runs' of CM$^{(m)}$ on a digital computer;
  \item letting the physical system that is modeled by CM$^{(m)}$
    physically evolve, and while it does that, measure physical
    observables which are predicted by the model.
  \end{itemize}
  When CM$^{(m)}$ is set up with a static architecture, one would
  expect that the execution
  dynamics follows mathematically  from the update operators
  $\sigma_v$ and $\alpha_v$ and the specific format of temporal
  progression, to which we alluded by the informal notation
  $\mathfrak{t} \mapsto \mbox{\sf next-}\mathfrak{t}$. However, local
  specifications of the mode of progression, which define how an
  individual observer state is updated (as in Equation
  \eqref{eStateUpdate}), may need to be augmented by rules for the
  global coordination of states that become updated 'simultaneously'
  or 'in parallel' or in 'threads' --- and what this mathematically
  means, concretely, is a question that must be answered for any
  concrete FC modeling formalism. In the digital computing world, a
  rich repertoire of formal methods has been developed to address
  concurrency in program execution \cite{Acetoetal07}. Rules for
  coordinating concurrent local dynamics would have to become a part
  of a model CM$^{(m)}$, except when the mode of progression simply
  specifies a global time, like $t \in \mathbb{R}$ in physics or $n
  \in \mathbb{N}$ in digital signal processing, which is shared by all
  variables in a model. 

  Besides a formal regulation of concurrency, further modeling
  elements need to be added to the concepts that we discussed so far,
  when it comes to dynamical architectures. In these architectures,
  the binding structure itself becomes time-dependent. In an intuitive
  graphical illustration (Figure \ref{figChronicleDynamics}) we
  collect some of the effects that we foresee in such self-modifying
  model architectures:

  \begin{itemize}
  \item Observers (and their chronicles) can join ({\bf a}) and split
    ({\bf b}), the latter giving rise to copying operations.
  \item Observers can terminate their existence or become created
    ({\bf c}). This effect might be formalized by the creation /
    delection of observers during execution; it could also be formally
    captured by declaring that an observer is 'instantiated' only when
    its activation is nonzero --- as indicated by the fading-out and
    fading-in in {\bf c}.
  \item Observers may bind into, and unbind from composite observers
    ({\bf d}). The central segment (broken orange outline) shows the
    temporary presence of a compound observer made from two primitive
    observers and one compound observer, which in turn is a binding of
    three primitive ones. The compound observers have an activation
    history of their own, which is not shown in the graphic. 
  \end{itemize}

  From these basic mechanisms one can obtain more complex effects. For
  instance, a sub-observer in a compound observer may split, resulting
  in an event where a compound observer spawns a copy instance of one
  of its member observers. All of these phenomena must be formally
  characterized. This can be done through adding structure-changing
  rules within CM$^{(m)}$, or (like it is the case for declarative
  programming languages) as a separate, external set of execution
  control rules.

\begin{figure}[htb]
  \center \includegraphics[width=5cm]{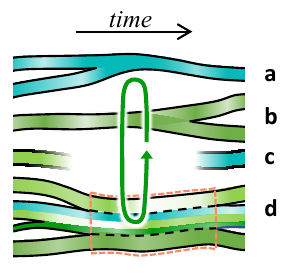}
  \caption{Some effects in self-modifying
    architectures. Coupling dynamics is symbolized by the circular
    arrow. Colored bands indicate chronicles of atomic observers, and their varying color intensity indicates varying
    activation.  For explanation see text.}
  \label{figChronicleDynamics}
\end{figure}

\end{description}

\subsection{Observers observe observers across modeling levels}

We repeat again that our prime motivation for interpreting variables
$v^{(m)}$ as observers is to establish computational models as models
of the underlying physical system PS, in a spirit of natural science
modeling or engineering. This let us consider the relation between two
subsequent models CM$^{(m)}$ and CM$^{(m-1)}$ as an observation
relation, where the source of what an observer $v^{(m)}$ observes lies
in the next-lower model CM$^{(m-1)}$ (or in the physical model PM for
observers $v^{(1)}$ in the machine interface model CM$^{(1)}$). These
sources are the observers $v^{(m-1)}$.  The level-transiting relations
between observers $v^{(m)}$ and $v^{(m-1)}$, marked as 'decoding' and
'encoding' in our general schema of computing system modeling (Figure
\ref{figModelStructure}), become interpreted in FC as observing and
being observed. Iterating these relations down the model hierarchy and
further through the physical model connects a higher-level model
CM$^{(m)}$ with its physical substrate.

At the same time we have used the 'observer' rationale also for the
within-model binding relations, which we (semi-)formalized through the
state update and activation update operators \eqref{eStateUpdate} and
\eqref{eActivationUpdateComplete}. To justify this secondary use of
the 'observer' concept we argued that in compilation hierarchies in
classical computer programming, the within-program definitions of
composite data structures and functions from simpler ones are of the
same kind as when a data structure or function, which is a primitive
in a higher-level programming language, becomes compiled into a
composite structure or process in a program written in a more
elementary programming language. The distinction of composition
hierarchies within versus between programs effectively blurs
when a higher-level programming language admits the direct inclusion
of lower-level code. In CS textbook theory the syntax of programming
languages is defined by context-free grammars. Such grammars can
switch to sub-grammars in their 'parse tree' representations of
computer programs, which provides a single mathematical format for
both within- and between-program compositionality.

In our FC proposal, we likewise use the same formal characterization
for within- and between-model observations.  Consider two models
CM$^{(m)}$ and CM$^{(m-1)}$. A designer of these two models connects
them by declaring observation relations between observers $v^{(m)}$ in
CM$^{(m)}$ and some observees $v_i^{(m-1)}$ in CM$^{(m-1)}$. Within a
model we referred to these observation relations as 'binding'. Across two models, we
will say that $v^{(m)}$ becomes \emph{defined} through the
$v_i^{(m-1)}$ that it observes. We adopt definitions from the previous
subsection as follows, assuming possibly different temporal
progression modes $\mathfrak{t}^{(m)}$ and $\mathfrak{t}^{(m-1)}$ at
the two modeling levels:
\begin{itemize}
\item Observers  $v^{(m)}$ in CM$^{(m)}$ are declared by the modeler
  to be either of type \emph{grounded} or of type \emph{free}. The
  type of $v^{(m)}$ remains fixed throughout the times when $v^{(m)}$
  is instantiated in an execution of CM$^{(m)}$.
\item Free observers  $v^{(m)}$ do not observe observers in
  CM$^{(m-1)}$.
\item For a grounded observer $v^{(m)}$, a set $\mathcal{G}_{v^{(m)}}$
  of observers $v_i^{(m)}$ in CM$^{(m-1)}$ is declared, called the
  \emph{grounding} of $v^{(m)}$. The grounding 
  $\mathcal{G}_{v^{(m)}}$ remains fixed during executions of
  CM$^{(m)}$.
\item For a grounded $v^{(m)}$, the modeler declares a state update
  rule, which essentially has the same form as the model-internal state
  update rule:
  \begin{equation}\label{eGroundedStateUpdate}
    \mathbf{s}_{v^{(m)}}(\mbox{\sf next-}\mathfrak{t}^{(m)}) =
    \tilde{\sigma}_{v^{(m)}}\left(\mathbf{s}_{v^{(m)}}(\mathfrak{t}^{(m)}),
    \left(\mathfrak{a}_{v^{(m-1)}}(\mathfrak{t}^{(m-1)})\right)_{v^{(m-1)} \in
      \mathcal{G}_{v^{(m)}}}\right).  
  \end{equation}
  This semi-formal schema is incomplete in that it does not explain
  how the temporal progression $\mathfrak{t}^{(m)}$ synchronizes with
  $\mathfrak{t}^{(m-1)}$. Additional specifications must be supplied by
  the modeler for concurrency bonds between
  $\mathfrak{t}^{(m)}$ and $\mathfrak{t}^{(m-1)}$. The idea in
  \eqref{eGroundedStateUpdate} is that $\mathfrak{t}^{(m)}$ and
  $\mathfrak{t}^{(m-1)}$ are 'simultaneous'. This needs to be worked
  out together with concrete FC modeling formalisms.
\item Our within-model binding relation lets the activation dynamics of
  $v^{(m)}$ be controlled top-down from master observers (Figure
  \ref{figCoupleBind}{\bf a}). In contrast, between models CM$^{(m)}$
  and CM$^{(m-1)}$ there is no top-down control. We thus have no
  equivalent of \eqref{eActivationUpdateComplete} here.   
  
\end{itemize}

We therefore have two hierarchies --- the within-model binding
hierarchy and the between-model grounding hierarchy --- which both
originate from the intuition of casting model variables as observers,
and whose state update rules mirror each other. This raises the
question, what are the deeper reasons for having two such similar
hierarchies, and what roles do they play in the modeling
game? This question is connected to fundamental issues in scientific
modeling and engineering practice at large, and cannot be answered
with a single answer. We explore some facets of this theme:

\begin{itemize}
\item Model hierarchies in the natural and neurosciences are often
  conceived as hierarchies of abstraction. When the cognitive
  reasoning dynamics of a brain is modeled in a classical AI
  formalism, all detail of neuron-level dynamics is collectivized,
  averaged, grouped, unified, simplified, etc.\ --- in short,
  'abstracted' (a word whose meaning at first sight seems clear like
  white light, but upon serious thinking splits into a rainbow
  spectrum of meanings). Stepping up
  the abstraction ladder through a  model hierarchy, 
  mechanical detail recedes from sight and functional
  insight comes closer. Information of some sort is lost, and of
  another sort is created ('information' being another of these
  meaning-full words). 

  There is a noteworthy difference between model abstraction
  hierarchies in the natural sciences, and hierarchies of AC
  programming languages and programs written in them. The latter
  admit \emph{exact} two-way translations, while the former do not. A
  program written in Python can be effectively translated to a
  program expressed in C, whose I/O functionality is exactly
  identical. Python is considered higher-level than
  C, and automated translation engines (compilers) exist. But it would
  equally be possible to translate C code to equivalent Python
  code, though this direction will hardly be used in practice.

  Let us take a closer look at translations from C to Python. There
  are two ways of doing this. The first is the easy one: the
  'fine-grained' data structures and functions of C code are 1-1
  recoded in Python. The resulting Python program \emph{simulates} all
  the minute operations in the C program. This is straightforward to
  automate, but it blows up the size of the program and runtime. The
  second way is much harder and essentially impossible to automate:
  find a \emph{compact} equivalent Python program. This means that the
  modeler has to \emph{discover} complex patterns within the C code
  that can be condensed into single Python data structures or
  commands. This creative, insightful upwards translation is sometimes
  called 'decompilation' in the CS world, with the additional
  connotation that this should be done in a more or less automated
  fashion. When decompilation is automated, the resulting higher-level
  code is often obscure and hard to make sense of by humans.

  Summarizing the intuitions and practice of compilation in classical
  programming: compilation from a higher-level language to a
  lower-level one is easy and automatic and turns conceptually more
  intuitive code into more efficient code that is closer to machine
  mechanics, while de-compilation from a lower-level language to a a
  conceptually better interpretable version is creative and hard or
  impossible to automate.

  The modeling hierarchies in the natural sciences are of the second,
  insightful, creative, bottom-up kind. There is no analogue of
  mechanical top-down compilation in scientific modeling hierarchies.

  Our FC proposal follows the natural science way of thinking about
  abstraction. Formulating a higher-level model CM$^{(m)}$ on the
  basis of CM$^{(m-1)}$, and linking it to that lower-level model by
  observation relations, reflects creative modeling insight. We do not
  expect that CM$^{(m)}$ can be distilled automatically from
  CM$^{(m-1)}$, nor CM$^{(m-1)}$ from CM$^{(m)}$.

\item A difference between 'abstraction' in FC model hierarchies
  versus model abstraction in the natural sciences lies in the
  scientific objectives that motivate the modeling efforts in the
  first place. In the sciences, the objective for creating models on
  all levels is to gain a deeply organized understanding of the
  physical system that lies underneath the finest-grained model at the
  bottom of the modeling hierarchy. In FC modeling, as well as in any
  cybernetic or algorithmic model hierarchy, the objective of
  increasing model abstraction is to connect a usecase or task that
  arises in an real-world task environment (TE in Figure
  \ref{figModelStructure}) to the physical computing system PS. Each
  step upwards from some CM$^{(m-1)}$ to CM$^{(m)}$ thus should help
  to 'morph' the computational model from the physical system model PM
  toward the task model TM (always referring to Figure
  \ref{figModelStructure}). In scientific model abstraction, 'more
  abstract' means a more globally organized understanding of a
  physical system PS, while in computational abstraction it means a
  more task-oriented understanding.

  This distinction between different modeling objectives is however
  most clear-cut in model hierarchies in physics and chemistry. In
  biology --- in neuroscience especially --- the target system may be
  regarded to have a purpose of its own, which the modeler wants to
  capture at the abstract high end of the modeling hierarchy. In
  computational terminology, there might be a 'task' which the
  biological system 'wants' to fulfil, or which it 'should'
  optimize. Such a 'task' might be quite globally stated as 'staying
  viable in a stochastic and possibly hostile environment'. This leads
  to modeling efforts geared at revealing robustness properties of
  biological systems, which is a prime theme in systems biology. In
  computational neuroscience, the 'task' of a brain may be formulated
  in more concrete terms, for instance as enabling the animal to
  predict the consequences of its actions, or as action planning, or
  even as being able to use language. To the extent that the overall
  modeling objective is task-oriented in some way, the nature of
  'abstraction' becomes similar to what we find in modeling computing
  systems.

\item A difference between abstraction in FC model hierarchies
  versus model abstraction in classical compilation hierarchies is
  connected to loss of information. Classical programs within a
  compilation hierarchy are functionally equivalent, while models in
  FC hierarchies will typically not be functionally equivalent. The
  higher-level temporal progression mode $\mathfrak{t}^{(m)}$ and the
  activation model $\mathfrak{a}^{(m)}$ may be coarser in some ways
  than $\mathfrak{t}^{(m-1)}$ and $\mathfrak{a}^{(m-1)}$. Furthermore,
  the modeler may simply ignore some of the substructures and
  subprocesses in CM$^{(m-1)}$ and only partially exploit
  CM$^{(m-1)}$, recruiting and re-interpreting only mechanisms in
  CM$^{(m-1)}$ that appear to be useful for the task.

\item The distinction between grounded and free observers $v^{(m)}$ is
  a matter of convenience in modeling practice. The execution of a
  model CM$^{(m)}$ must not lead to innovative dynamics that cannot be
  reduced to the execution dynamics of CM$^{(m-1)}$. The modeler may
  decide to define a free observer $v^{(m)}$ only within CM$^{(m)}$,
  by declaring it as a compound observer that binds some components
  $v_i^{(m)}$.  The level-$m$ observers in
  ${\mathcal{B}^\downarrow}^\ast_{v^{(m)}}(\mathfrak{t})$, that is all
  the members in the component tree underneath $v^{(m)}$, must
  ultimately be grounded: every subcomponent path
  $v^{(m)} \rhd_{\mathfrak{t}} v_1^{(m)} \rhd_{\mathfrak{t}} v_2^{(m)}
  \rhd_{\mathfrak{t}} \ldots$ must ultimately lead to grounded
  observers. Thus, an observer $v^{(m)}$ that is declared by the
  modeler without explicit grounding in CM$^{(m-1)}$, could also
  equivalently be declared via grounding through all its grounded
  (transitive) 
  components.

  This situation is analogue to a common practice in machine
  learning. In graphical models \cite{Jordan04} one frequently
  introduces 'hidden' variables, which are random variables whose
  values are not directly observable in the modeled target domain, but
  whose distributions are ultimately conditioned on observable
  variables.

\item Abstraction from CM$^{(m-1)}$ to CM$^{(m)}$ may entail a loss of
  information through coarsening and omitting subsystems. But at the
  same time, the creative selection, re-combination and
  re-interpretation of subprocesses in CM$^{(m-1)}$, which moves
  CM$^{(m)}$ closer to the conceptual framing of the task model TM,
  can be regarded as adding 'knowledge' that stems from the modeler's
  prior insight in the task conditions. This added influx of modeler's
  knowledge can intuitively also be called 'information', although
  this sort of information is certainly not Shannon information in the
  sense of information theory. The mathematical format of this sort of
  information is unclear, and making it formal will depend on the
  formal workout of a concrete FC formalism. An analogue is the
  'Bayesian prior' known in Bayesian statistics.

\item The commutativity conditions that we pointed out in Section
  \ref{secStrucPhysTheory} must be satisfied in FC model hierarchies.

\item One natural way of abstracting  CM$^{(m-1)}$ to CM$^{(m)}$ is to
  \begin{itemize}
  \item set the variables $\mathcal{V}^{(m)}$ in CM$^{(m)}$ equal to
    (copies of) the variables in $\mathcal{V}^{(m-1)}$ that are
    highest-level compounds in CM$^{(m-1)}$, that is to re-use only the
    most complex observers from CM$^{(m-1)}$,
  \item and then find suitable temporal progression modes, state
    representations and activation laws between those compound
    observers that let the diagrams of CM$^{(m-1)}$ and CM$^{(m)}$
    commute.
  \end{itemize}

\end{itemize}

\subsection{Some general observations on observations in FC
  modeling}\label{secGenObs}

\begin{figure}[htb]
  \center
  \includegraphics[width=14cm]{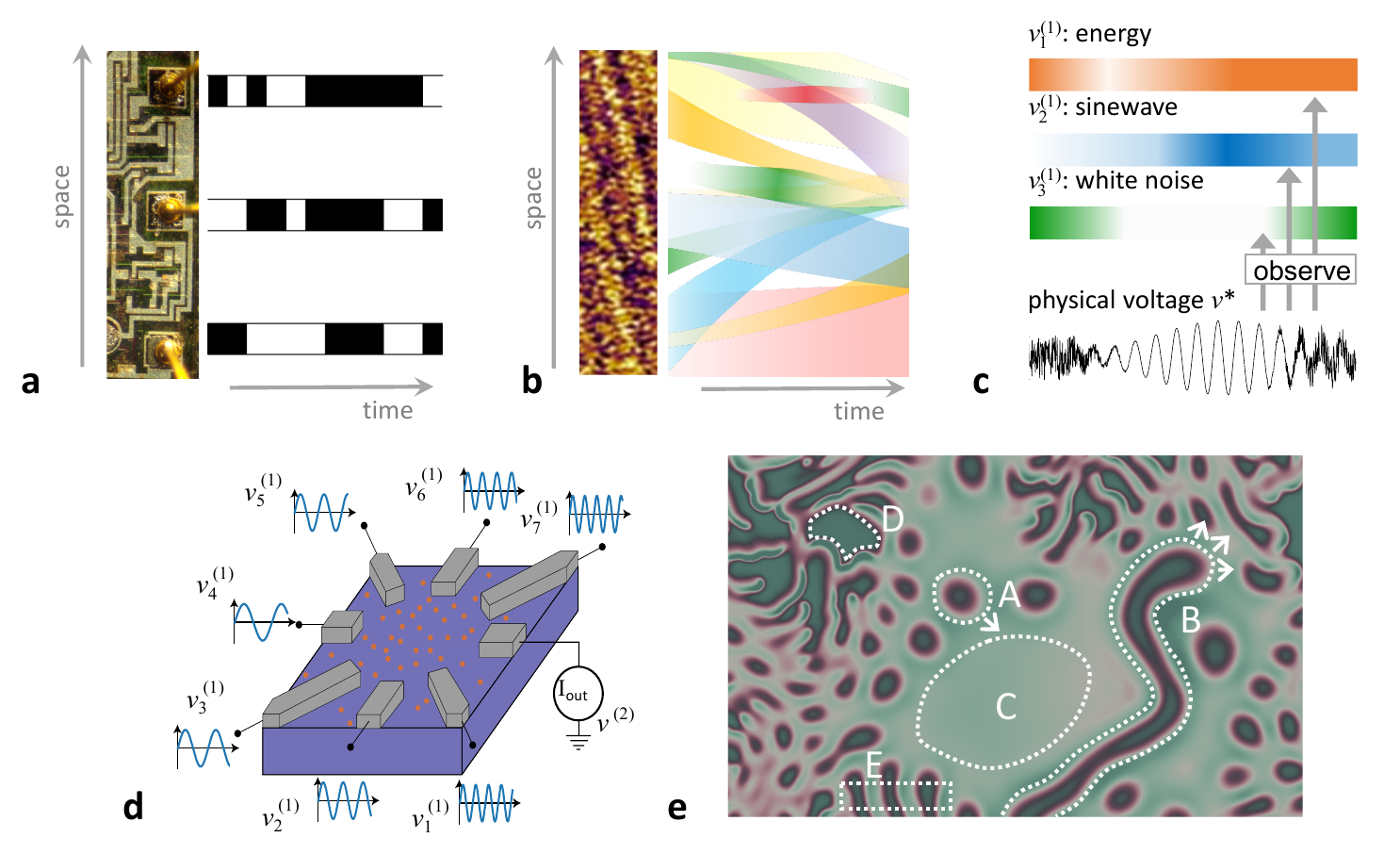}
  \caption{Highlighting some aspects of observers. For explanation see
    text.}
  \label{figOnObservations}
\end{figure}

In Figure \ref{figOnObservations} we illustrate some facts and
examples about observations that are pertinent to building model
abstraction hierarchies in an FC spirit.

\begin{description}
\item[Localizing bit signals ({\bf a}).] An observer --- whether a
  voltmeter or a human --- does not observe the observed physical
  system at large but focuses on a segment of it. Specifying an
  observer includes a specification of its source. Often this will
  boil down to specifying a physical location.  The binary bit
  switching signals in a digital microchip can be picked up at
  pointlike, non-moving localizations.

\item[Observables can be spatially extended ({\bf b}).] In physical
  substrates (microphotography shows the material from Figure
  \ref{figDomainWalls}) one can observe objects or phenomena that are
  spatially extended, geometrically time-varying, and moving ---
  fields, wave fronts, particles and more. An observer declaration
  must include means to identify, localize, and track such objects.

\item[There are unlimited ways to define observers ({\bf c}).]  The
  graphic shows records from three observers $v_1^{(1)}$, $v_2^{(1)}$,
  $v_3^{(1)}$ that observe the voltage $v^\ast$ of an electronic
  contact point, where here $v^\ast$ is a variable in the physical
  model PM. Their activations (indicated by color intensity) respond to
  the short-term averaged signal energy, a specific sine frequency
  response, and the white noise component, respectively. In an FC
  model there can be an infinity of observers of different quality at
  every level of modeling abstraction, and even an infinity of
  observers of different quality that observe the same source.

\item[A complex physical observer ({\bf d}).] Observers can have
  internal state and memory, enabling them to integrate and filter
  information over time. The drawing (adapted from
  \citeA{RuizEuleretal20}) shows a schematic of a dopant network
  processing unit (DNPU) \cite{Chenetal20}, an unconventional
  nanoscale device developed in the lab of co-author WvdW. It consists
  of a doped silicon well which here is contacted by seven input and
  one output electrode. DNPUs exhibit strongly nonlinear charge
  transport behaviour between the electrodes. They can implement
  different computational models, including neural circuits. The
  drawing shows an experiment where different input voltage signals
  lead to a nonlinear current response in the output. A neural network
  was trained to model this highly nonlinear 7-input, 1-output
  function. Using this model, modular DNPU architectures were
  mathematically optimized to yield very compact signal processing and
  pattern recognition systems, including handwritten digit recognition
  \cite{RuizEuleretal20b}, which were then tested with physical DNPUs.

  This neural network model of a DNPU can be formally cast as an
  observer of the seven inputs $v^{(1)}_1 - v^{(1)}_7$ together. The
  activation signal $v^{(2)}$ of the trained model output neuron yiels
  the chronicle of this observer.

\item[Observing spatiotemporal patterns ({\bf e}).] This graphic
  (background image created with the fabulous online tool of
  \citeA{Sims23}) shows a snapshot from an evolving
  chemical reaction-diffusion system. We might consider as observable
  phenomena, for example, moving solitons (A), moving and growing
  filaments (B), neutral ground state areas (C), activated areas (D),
  periodic patterns (E).  Arbitrarily more pattern categories can be
  defined.
\end{description}

\section{Algorithmic models seen as fluent   models} \label{secAlgAsFluent}

Digital computers are physical machines, and our FC proposal aims at
modeling general physical computing systems. Thus it should be
possible to re-cast the workings of digital or other
symbolic-algorithmic machines in the FC modeling framework. How this
can be done depends in its details on the specific AC model that one
wishes to re-formulate as an FC model. Here we explain in general terms
the main steps needed to transform classical AC models into FC models.  

The main hurdle is that AC state variables
$v_{\mbox{\tiny \sf AC}}^{(m)}$ take arbitrarily nested symbol
structures as values, while FC state variables
$v_{\mbox{\tiny \sf FC}}^{(m)}$ have 'scalar' activation values
$\mathfrak{a}_{v_{\mbox{\tiny \sf FC}}^{(m)}}$. At low modeling levels
CM$^{(m)}$ close to the physical hardware, by default we think of such
activations as non-negative real numbers $a \in \mathbb{R}^{\geq 0}$;
at more abstract higher modeling levels we will often want to use
reduced-precision versions, like natural numbers or suitably defined
fuzzy intervals of $\mathbb{R}^{\geq 0}$. But in any case, an
activation value should be in some sense 'one-dimensional' and not
internally structured.

We consider some fixed AC modeling level CM$^{(m)}$ and will again
drop the superscript $\cdot^{(m)}$ for better readability.  For
simplicity let us assume that the AC model uses global discrete update
steps $n \in \mathbb{N}$. Consider the situation that after step $n$,
an AC variable $v_{\mbox{\tiny \sf AC}}$ has been assigned the nested
symbolic data structure $D$ as its value, that is
$v_{\mbox{\tiny \sf AC}}(n) = D$.  In order to re-formulate this fact
in terms of unstructured scalar activations, we can mirror the
compositional structure of $D$ in an analogue compositional binding
structure of a composite observer $v_{\mbox{\tiny \sf FC}}$, and
encode the symbols at the innermost nesting level in $D$ by what is
known as 'one-hot' encoding in machine learning.

\begin{figure}[htb]
  \center
  \includegraphics[width=14cm]{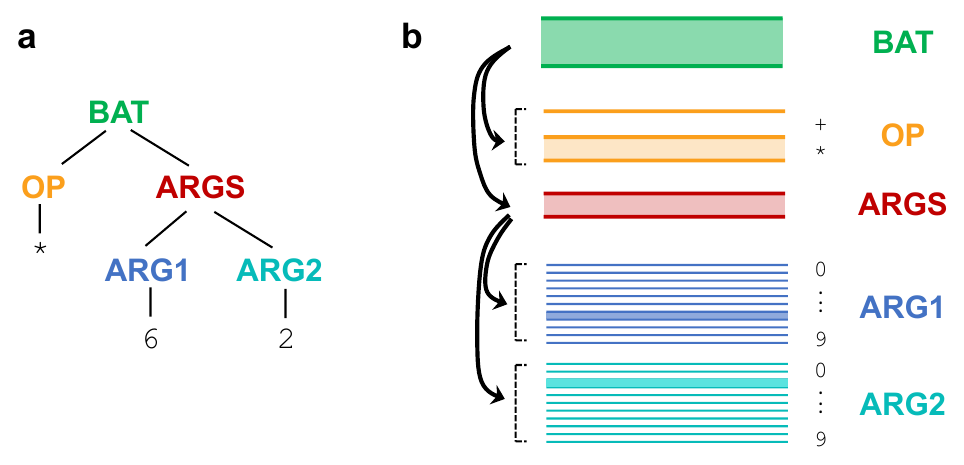}
  \caption{{\bf a:} Parse tree for the binary arithmetic term {\tt
      6*2}. {\bf b:} Its re-formulation in an observer binding
    structure. For explanation see text. }
  \label{figParseTree}
\end{figure}

This is best explained with an elementary example. Consider a symbolic
term $D = $ {\tt 6*2}. AC theorists would bring the hierarchical
nesting structure of $D$ to the front by drawing its \emph{parse
  tree} according to the syntax grammar of the used AC modeling
formalism (Figure \ref{figParseTree}{\bf a}). This could be done by
introducing variables {\sf BAT} ('binary arithmetic term'), {\sf OP}
('operator'), {\sf ARGS} ('arguments'), {\sf ARG1} ('first argument'),
{\sf ARG2} ('second argument') for the grammatical substructures, and
\emph{terminal symbols} {\tt +}, {\tt *}  and {\tt 0}, $\ldots$, {\tt
  9} for the possible operators and argument values.

In an FC re-formulation, the grammatical variables {\sf BAT}, {\sf OP},
{\sf ARGS}, {\sf ARG1}, {\sf ARG2} as well as the terminal symbols
{\tt +}, {\tt *}, {\tt 0}, $\ldots$, {\tt 9} become observers, with a
binding structure identical to the parse tree structure (Figure
\ref{figParseTree}{\bf b}). The top-level observer {\sf BAT} is bound
from two sub-observers {\sf OP} and {\sf ARGS}; the observer {\sf OP}
is a compound of two atomic observers {\tt +} and {\tt
  *}; the {\sf ARGS} observer is obtained by binding together {\sf
  ARG1} with {\sf ARG2}; and these latter ones are each composites of
ten atomic observers {\tt 0}, $\ldots$, {\tt 9}.

The fact that the specific value of $v_{\mbox{\tiny \sf AC}}$ at time
$n$ is {\tt 6*2} (and not, for instance, {\tt 3+7}) can be encoded,
for instance, in the FC version by setting the activation of the {\tt
  *} observer to 1 and of {\tt +} to 0, and similarly setting the
activations of {\tt 6} and {\tt 2} inside {\sf ARG1} and {\sf ARG2} to
1, while the other atomic digit observers have zero activation. The
unit activations are marked in Figure \ref{figParseTree}{\bf b} by
color filling. The
higher-level observers {\sf BAT} and {\sf ARGS} also get unit
activations.  More formally, in the FC version we would use the
discrete temporal progression $\mathfrak{t} = n \in \mathbb{N}$ and
set $\mathfrak{a}_{\scriptsize \mbox{\sf BAT}}(n) = 1$,
$\ldots$,  $\mathfrak{a}_{{\scriptsize \mbox{\tt 2}}\mbox{\scriptsize
    -in-}{\scriptsize \mbox{\sf ARG2}}}(n) = 1$.

This 'one-hot' encoding scheme for re-formulating compound AC variable
values as 0/1 activation patterns in compound FC observers is not the
only option. For instance, one could alternatively encode the 10
different digit values of {\sf ARG1} or {\sf ARG2} in an AC model by
ten different activation levels of atomic observers {\sf ARG1} or {\sf
  ARG2} in an FC model.

In an AC computation process, a model variable
$v_{\mbox{\tiny \sf AC}}$ need not have a defined value at all times
during the computation. In an FC re-modeling, this can be accomodated
by setting the activations of the mirror observer
$v_{\mbox{\tiny \sf FC}}$ to zero at times $n$ when it has no defined
value.  Furthermore, new variables can be dynamically created or
deleted during an AC computation. This would be mirrored in an FC
setting by dynamical model architectures.

\begin{figure}[htb]
  \center
  \includegraphics[width=14cm]{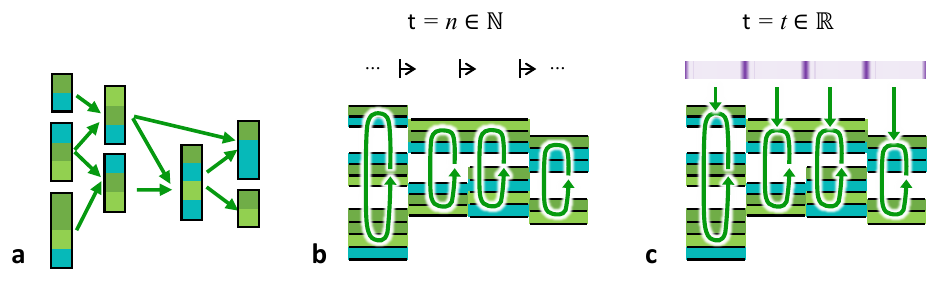}
  \caption{An algorithmic computation ({\bf a}, extracted from Figure
    \ref{figAlgCyb}) re-cast as a fluent computation in a discrete
    ({\bf b}) and a continuous mode of progression ({\bf c}). The
    hierarchical sub-structuring of observers, illustrated in Figure
    \ref{figParseTree}, is not shown in {\bf b} and {\bf c}. For
    explanation see text. }
  \label{figReinterpret}
\end{figure}

Figure \ref{figReinterpret} illustrates in a schematic way how a
sequence of threaded AC operations can be re-cast as a fluent
computation with a discrete mode of progression (panel {\bf b}) or a
continuous one (panel {\bf c}). In the latter case, the operations of
the digital clock, which is needed to coordinate the within- and
between-thread data transformations steps, can be accomodated by a
'clock observer' $v_{\mbox{\scriptsize tick}}$, whose activation
history is shown in the violet bar on top of the graphic. This
observer is blind - it observes no other observers. It is an
autonomous activity oscillator. Using continuous time
$t \in \mathbb{R}$, its state and activation updates (Equations
\ref{eStateUpdate} and \ref{eActivationUpdateComplete}) reduce to
$\dot{\mathbf s}_{v_{\mbox{\tiny tick}}} = \sigma_{v_{\mbox{\tiny
      tick}}}({\mathbf s}_{v_{\mbox{\tiny tick}}})$ and
$\mathfrak{a}_{v_{\mbox{\tiny tick}}}(t) = \alpha_{v_{\mbox{\tiny
      tick}}}({\mathbf s}_{v_{\mbox{\tiny tick}}}(t))$. The format of
the state update law allows the formulation of any desired
oscillator dynamics. The clock observer $v_{\mbox{\scriptsize tick}}$
would become a component of every other observer in the model
(indicated by vertical green arrows in Figure \ref{figReinterpret}{\bf
c}).

Of course all of this is only a superficial sketch, which would need
to be specified in detail according to the specific formalism that one
uses for an FC model.

\section{Summary, discussion and outlook}\label{secConclusion}

Research in neuromorphic and other unconventional kinds of computing
is thriving, but still lacking a unifying theory grounding. We propose
to anchor such a theory in three ideas: viewing information processing
as a dynamical system (adopted from the cybernetic paradigm),
organizing these dynamics in hierarchical binding compounds (adopted
from the algorithmic paradigm), and ground theory abstraction in
hierarchies of formal observers (following physics). 

The core of our FC proposal is to construe formal computational
objects as abstract observers of physical phenomena in the material
substrate of a computing system. This reverses of Turing's
view. Turing conceived of discrete symbolic structures and their
transformations as a model of rational, symbolic reasoning. This
perspective makes AC models easily ``thinkable'' for humans, because
there is a pre-established harmony between the reasoning of a
well-trained programmer and the structure of computer programs. The
downside is that the discrete symbolic structures and discrete-step
mechanisms of digital computer programs (or other AC models) must be
\emph{imposed} on the physical hardware. Only hardware systems which
are engineered around discrete switching dynamics can serve as
physical bases for algorithmic computing. In our reversed perspective,
any observable physical phenomenon is a candidate carrier for
computing. The downside here is that FC models cannot be directly
mapped to rational, discretely rule-based, symbolic human reasoning.
But they might turn out to be readily mappable to other modalities of
human cognitive processing --- associative thinking, sensori-motor
perception-action dynamics, continual adaptation and learning, graded
concepts and other cognitive phenomena that indeed happen in our
brains. A general theory of physical computing --- whether it is
shaped after our FC proposal or in some other way --- should do good
service as a mathematical tool for modeling processes in biological
brains.

Surely there are many ways how the FC ideas, which at present we can
only outline in a schematic manner,
can be tied together in mathematical detail. There are a number of
nontrivial tasks that must be solved in order to obtain a full
mathematical workout, among them the following:

\begin{itemize}
\item Define a hierarchally organized systems of temporal progressions
  $\mathfrak{t}$, presumably starting from the standard continuous
  timeline model $t \in \mathbb{R}$ used in physical modeling, and
  leading to the entirely de-materialized progression of successive
  logical inference steps that is used in the Turing machine
  model. Similarly, an abstraction hierarchy of activation models
  $\mathfrak{a}$ must be established.
  
\item Define binding operators for observers. In algorithmic modeling
  formalisms, the format of all data structures can ultimately be
  reduced to writing them as trees, or equivalently, nested 
  lists. For an FC formalism one may want to consider other options
  too, for instance grouping sub-observers in unstructured sets, or
  declaring weighted connections between them.

\item Define mathematical formats for the state update functions
  $\sigma_v$ and activation functions $\alpha_v$. These have to agree
  with the chosen models for temporal progression and activation
  values. Furthermore, in order to obtain 'executable' models
  CM$^{(m)}$, global coordination mechanisms must be defined, which
  regulate the interaction of subprocesses. This will likely require
  the introduction of geometrical space concepts in which observers
  can be localized.

\item In order to specify commuting diagram conditions one needs
  measures of similarity between two versions of observers and their
  activations, which are determined by within-level-$m$
  transformations and cross-level determination pathways (like the two
  $x_2^{(1)}$ versions in Figure \ref{figCommute}).

\item In this article we have not started to investigate how in an FC
  modeling scenario the task model TM (recall Figure
  \ref{figModelStructure}) should be devised. From a user perspective,
  many non-digital computing systems will behave quite differently
  from digital computers. They may not be programmable in the
  customary sense but need to be trained; they may execute
  stochastically; they will operate continuously and may not be
  rebootable or even not re-startable; their output signals may be
  neither numerical nor symbolic but physical of some sort which needs
  other mathematical formats to be appropriately captured (for
  instance through topological objects in dynamical systems or
  probability distributions \cite{Jaeger21a}). The logic formalisms
  used in AC for task modeling are not immediately suited to capture
  such aspects, and new mathematical modeling perspectives may have to
  be found.  In the historical development of AC, the problem of
  formalizing tasks and connecting them to the procedural mechanics of
  computational models through program verification techniques came
  into view only after digital computers became widely used. We
  would expect that developing TM modeling formalisms for FC can
  likewise be postponed.
  
\end{itemize}

All of this together looks somewhat fearsome. However, we have good
hopes that we can find the right mathematical abstractions that make
the challenge is less daunting and the resulting FC formalism less
complex than how it might appear at first sight. The main idea is to
express the dynamical effects of state update operators, in a spirit
of formal algebra, in terms of changes to the binding structure
itself. This line of work has been pursued by H.J.\ for several years
and is making good progress, but we cannot promise a publishable
ready-to-use yet.

\begin{figure}[htb]
  \center
  \includegraphics[width=14cm]{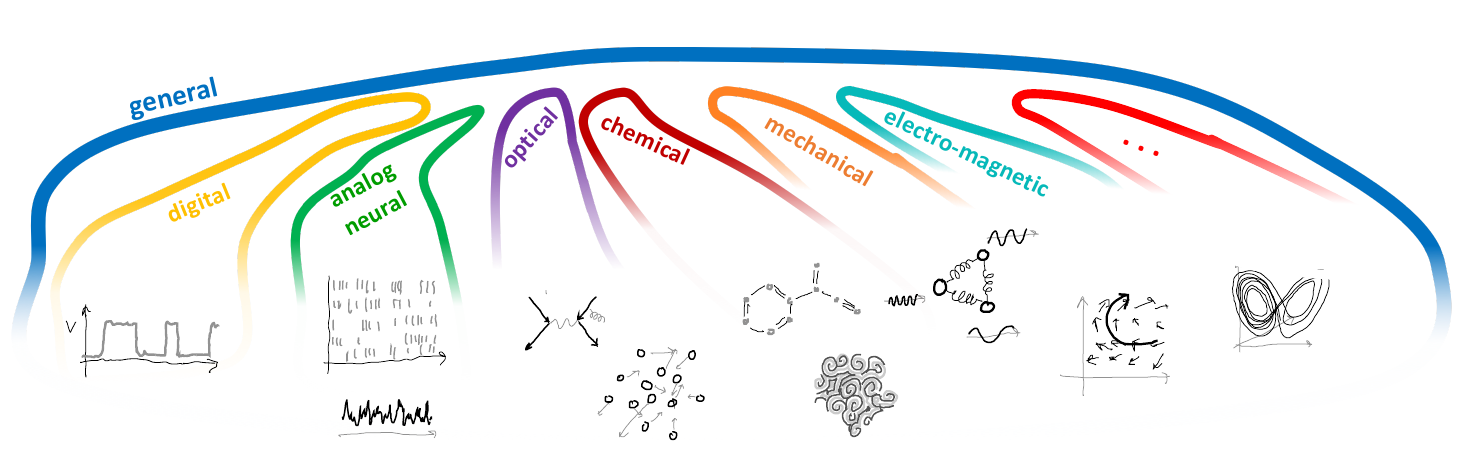}
  \caption{A general theory of physical computing systems would comprise
    existing theories as special cases.}
  \label{figGeneralTheory}
\end{figure}

Let us briefly reflect on what a worked-out general theory of physical
computing systems (whether of an FC sort or other) could give us:

\begin{description}
  \item[Unified terminology:] Researchers who today are using different
conceptual framings and terminologies could formulate their insights
in the shared scientific language of a GFT.
\item[Model translations:] In each of the currently used modeling
  frameworks we could identify what is essential for computing, and
  translate this into other languages and models. A role model for
  such substance-identifying and substance-preserving formalism
  translations are the meta-models of \emph{universal logic} that are
  developed in theoretical computer science. A readable introduction
  to this advanced field of mathematical logic is given in
  \citeA{Rabe08}; \citeA{Rabe17} gives a recent synopsis. The tools
  from universal logic allow logicians to identify the essential
  logical (i.e., proof-enabling) ideas that mathematicians use in
  their varied subbranches of mathematics (algebra, calculus,
  probability and all others) and to translate them across different
  mathematical terminologies, formalisms, and axiomatic
  frameworks. Similarly, a general theory of physical computing
  systems may enable us to compare and interrelate computational
  mechansism that arise from seemingly quite different physical
  phenomena in different sorts of computational materials (Figure
  \ref{figGeneralTheory}). 
\item[Physical phenomena translations:] A diversity of nonlinear
  micro- or nanoscale effects is currently being explored for
  computational exploits in the materials sciences. At present we do
  not have conceptual tools (let alone rigorous mathematical
  definitions) of how to identify, across this sparkling spectrum of
  physical phenomena, those invariances that are constitutive for
  computational function. All we can see today is a small number of
  basic \emph{dynamical} similarities (like the characterization of
  hysteretic or oscillatory dynamics), \emph{statistical} equivalences
  (like types of phase transitions) or \emph{structural-geometrical}
  analogies (like boundary formation or filament growing). But we do
  not yet understand how such invariances may become coupled into
  computationally functional complex bundles of mechanisms, and very
  likely we are still blind to many types of physical phenomenal
  invariances that are important for general computing functions.
\item[Simulation models:] A worked-out model hierarchy for computing
  systems should be concrete enough to enable the design of generic
  simulation tools, runnable on digital computers like all formal
  models in the natural sciences. These simulations would allow us to
  experimentally explore generic computational systems that are
  formulated independent of specific physical substrates or modalities
  (optical, electronic, mechanic etc.).
\item[General computing as a natural science:] A GFT would enable
  us to to generate falsifiable hypotheses (testable in generic
  simulation or specific physical experiments) and predict
  computational phenomena that have not yet been observed in reality.
\item[Research transfer between natural and engineered systems.]  Many
  formal models that are specializations of a GFT
  (Figure \ref{figGeneralTheory}) may be dually applicable to natural
  and engineered systems. This is clear for analog models of neural
  systems that are conceived at the interface between computational
  neuroscience and neuromorphic engineering. There are also digital
  models of biological neural systems, for instance the classical
  McCullogh-Pitts model \cite{McCullochPitts43}.
\item[System comparison and classification:] In analogy to the deeply
  insightful classification system for AC algorithms offered by
  computational complexity theory \cite{FortnowHomer03}, a general
  theory of physical computing systems could help us to define and
  compare different sorts of computing systems, classified according
  to their power and cost. In computational complexity theory
  different sorts of cost have been defined, like runtime or memory
  consumption. For general computing systems, new sorts of cost may be
  identified and formally characterized, for instance relating to
  energy consumption, physical endurance and replicability, or other
  criteria which are outside the classical AC perspective but come
  into view when physical properties of the underlying hardware
  substrates become essential modeling targets \cite{Blakey17}.
 
\item[Transcending computing:] Finally, the new formal modeling tools
  and their underlying conceptual intuitions may extend our scientific
  methods repertoir to describe and analyse general complex systems
  which are not primarily seen as computing systems, in particular
  biological and social systems.  
\end{description}

We hope that this article gives useful
orientation for theory builders who, like ourselves, are searching for
the key to unlock the richness of material physics at large for
engineering neuromorphic and other unconventional computing systems.

\vspace{0.5cm}

\noindent {\bf Acknowledgements} H.J. acknowledges financial support
from the European Horizon 2020 projects
\href{https://memscales.eu/}{\emph{Memory technologies with
    multi-scale time constants for neuromorphic architectures}} (grant
Nr.\ 871371) and
\href{https://postdigital.astonphotonics.uk/}{\emph{Post-Digital}}
(grant Nr.\ 860360). W.G.v.d.W acknowledges financial support from the
HYBRAIN project funded by the European Union's Horizon Europe research
and innovation programme under Grant Agreement No 101046878 and the
Deutsche Forschungsgemeinschaft (DFG, German Research Foundation)
through project 433682494 -- SFB 1459. B.N. acknowledges funding  from the European Union’s Horizon 2020 ETN programme Materials for Neuromorphic Circuits (MANIC) under the Marie Skłodowska-Curie grant agreement No 861153. Financial support by the
Groningen Cognitive Systems and Materials Center (CogniGron) and the
Ubbo Emmius Foundation of the University of Groningen is gratefully
acknowledged. Finally, we want to thank the reviewers of the
short journal version \cite{JaegerNohedaVdWiel23a} of this arXiv article for
their extensive, insightful and constructive feedback. In this long
version we tried to  accomodate all of their valuable
suggestions.

\bibliography{./referencesPerspective4Arxiv.bib} 

\end{document}